\documentclass[a4paper,11pt]{article}
\usepackage[DIV12]{typearea}
\usepackage{longtable}
\usepackage{booktabs,colortbl}
\usepackage{multirow}
\usepackage{amsmath}
\usepackage{amssymb}
\usepackage{bbm}
\usepackage[utf8]{inputenc}
\usepackage{pdflscape}
\usepackage{xspace}
\usepackage[caption=false]{subfig}
\usepackage{nicefrac}
\usepackage{graphicx}
\usepackage{cite}  
\usepackage[small]{caption2}
\usepackage{mathtools}
\usepackage{nccfoots}
\usepackage{tikz}
\usetikzlibrary{calc,positioning}

\usepackage[all,cmtip]{xy}
\newlength{\xywd}
\newcommand{\xyrightarrow}[2][]{%
  \sbox{0}{$\scriptstyle#1$}%
  \xywd=\wd0
  \sbox{0}{$\scriptstyle#2$}%
  \ifdim\wd0>\xywd \xywd=\wd0 \fi
  \xymatrix@C\dimexpr\xywd+1em\relax{{}\ar[r]^{#2}_{#1}&{}}%
}

\newcommand{\rep}[1]{\ensuremath\boldsymbol{#1}}
\newcommand{\crep}[1]{\ensuremath\bar{\boldsymbol{#1}}}
\newcommand{\Z}[1]{\ensuremath{\mathbbm{Z}_{#1}}} 

\newcommand{\SL}[1]{\ensuremath{\mathrm{SL}(#1)}}
\newcommand{\GL}[1]{\ensuremath{\mathrm{GL}(#1)}}

\newcommand{\e}{\mathrm{e}}
\newcommand{\I}{\mathrm{i}}
\newcommand{\Id}{\mathbbm{1}}

\newcommand{\CP}{\ensuremath{\mathcal{CP}}\xspace}
\newcommand{\x}{\ensuremath{\times}}
\newcommand{\vev}[1]{\ensuremath{\langle{#1}\rangle}}

\usepackage{xcolor}
\definecolor{darkgreen}{HTML}{109930}
\definecolor{pink}{rgb}{0.858, 0.188, 0.478}

\advance \headheight by 3.0truept       
\usepackage[pdftex,hidelinks,colorlinks=true,citecolor=darkgreen]{hyperref}
\hypersetup{
    pdftitle = {Top-Down Anatomy of Flavor Symmetry Breakdown},
    pdfauthor = {Baur, Nilles, Ramos-Sanchez, Trautner, Vaudrevange}
}

\begin{document}

\begin{titlepage}

\begin{flushright}
\normalsize{LMU-ASC 47/21, TUM-HEP 1377/21}
\end{flushright}

\vspace*{1.0cm}

\begin{center}
{\Large\textbf{\boldmath Top-Down Anatomy of Flavor Symmetry Breakdown  \unboldmath}}

\vspace{1cm}
\textbf{Alexander~Baur}$^{a,b}$,
\textbf{Hans~Peter~Nilles}$^{c,d}$,
\textbf{Sa\'ul Ramos--S\'anchez}$^{b}$,\\
\textbf{Andreas Trautner}$^{e}$, and 
\textbf{Patrick~K.S.~Vaudrevange}$^{a}$
\Footnote{*}{%
\href{mailto:alexander.baur@tum.de;nilles@th.physik.uni-bonn.de;ramos@fisica.unam.mx;trautner@mpi-hd.mpg.de;patrick.vaudrevange@tum.de}{\tt Electronic addresses} 
}
\\[5mm]
\textit{$^a$\small Physik Department, Technische Universit\"at M\"unchen,\\ James-Franck-Stra\ss e 1, 85748 Garching, Germany}
\\[2mm]
\textit{$^b$\small Instituto de F\'isica, Universidad Nacional Aut\'onoma de M\'exico,\\ POB 20-364, Cd.Mx. 01000, M\'exico}
\\[2mm]
\textit{$^c$\small Bethe Center for Theoretical Physics and Physikalisches Institut der Universit\"at Bonn,\\ Nussallee 12, 53115 Bonn, Germany}
\\[2mm]
\textit{$^d$\small Arnold Sommerfeld Center for Theoretical Physics, 
Ludwigs-Maximilians-Universit\"at M\"unchen,\\
Theresienstra\ss e 37, 80333 M\"unchen, Germany}
\\[2mm]
\textit{$^e$\small Max-Planck-Institut f\"ur Kernphysik, \\ Saupfercheckweg 1, 69117 Heidelberg, Germany}
\end{center}

\vspace{1cm}

\vspace*{1.0cm}

\begin{abstract}
A top-down approach to the flavor puzzle leads to eclectic
flavor groups which include modular and traditional flavor
symmetries. Based on examples of semirealistic $\mathbbm{T}^2/\Z{3}$ 
orbifold compactifications of heterotic string theory,
we discuss the breakdown patterns of the 
eclectic flavor group via the interplay of vacuum expectation 
values (vevs) of moduli and flavon fields. 
This leads to an attractive flavor scheme with various possibilities to obtain 
``flavor hierarchies'' through the alignment of these vevs. 
Despite the fact that the top-down approach gives strong 
restrictions for bottom-up flavor model building, it seems to be well 
suited to provide a realistic flavor pattern for quarks and 
leptons.
\end{abstract}

\end{titlepage}

\newpage

\section{Introduction}

In his influential work~\cite{Feruglio:2017spp}, Feruglio considered finite modular groups
for flavor model building. This proposal has triggered intense activity in the construction 
of bottom-up (BU) models of quark and lepton interactions based on various finite modular 
groups~\cite{Criado:2018thu,Kobayashi:2018vbk,deAnda:2018ecu,Penedo:2018nmg,Ding:2019zxk,Ding:2020yen} 
(see e.g. the review ref.~\cite{Feruglio:2019ybq} for a complete set of references and 
further details on the BU approach). One successful result of these constructions is 
that they can provide  good fits to data by typically requiring a smaller number of 
free parameters than in models endowed with traditional flavor symmetries. However, the predictability 
of such BU constructions may be challenged through the appearance of uncontrollable terms 
in the K\"ahler potential~\cite{Chen:2019ewa}. In the BU approach there are thus many working 
models with some degree of predictability but we still lack a baseline theory
coming from an underlying fundamental principle.

To address this problem, there has been extensive work devoted toward top-down (TD) derivations 
of modular flavor symmetries from ultraviolet complete models based on string theory~\cite{Kobayashi:2018rad, Baur:2019kwi}. 
Apart from heterotic orbifold compactifications~\cite{Baur:2019kwi, Baur:2019iai}, 
TD models include scenarios based on compactifications on toroidal orientifolds~\cite{Kobayashi:2020hoc} and 
magnetized tori~\cite{Kobayashi:2018bff, Kariyazono:2019ehj,Ohki:2020bpo,Almumin:2021fbk}.
TD constructions typically give strong restrictions on the allowed symmetries 
and the particle spectrum, and may also allow one to constrain (or even compute) the 
otherwise free terms of the K\"ahler potential~\cite{Nilles:2020kgo,Chen:2021prl,Baur:2021mtl,Almumin:2021fbk}.
Furthermore, TD models naturally include unification of flavor with \CP-like 
transformations~\cite{Baur:2019kwi, Baur:2019iai,Novichkov:2019sqv,Kobayashi:2019uyt}.
This unification becomes even more transparent in models where the modular group
is extended to its metaplectic~\cite{Liu:2020msy,Yao:2020zml,Almumin:2021fbk}
or symplectic cover~\cite{Ding:2020zxw,Ding:2021iqp,Baur:2020yjl,Nilles:2021glx,Ishiguro:2020nuf}
as recently discussed both in the BU and TD approaches.

In the present paper, we concentrate on the TD approach. This approach to the flavor problem
leads to a holistic view that necessarily encompasses 
all available kinds of discrete flavor symmetries. It thus has to include traditional flavor 
symmetries and $R$-symmetries as well as finite modular flavor symmetries and their associated 
\CP transformations~\cite{Baur:2019kwi}. These symmetries 
reflect the symmetries of the underlying UV complete theory, which we consider here
in the framework of heterotic strings with compactifications with elliptic fibrations. The discrete
flavor symmetries can be derived in full generality as the outer automorphisms of the Narain space 
group~\cite{Baur:2019kwi,Baur:2019iai}. This leads to the concept of the eclectic flavor group 
\cite{Nilles:2020nnc, Nilles:2020kgo, Nilles:2020tdp,Baur:2021mtl}
as a multiplicative closure of all flavor symmetries. The eclectic flavor group is the maximal 
possible flavor group, but it is only partially linearly realized. The linearly realized flavor 
subgroup is nonuniversal in moduli space, a property that leads to the concept of 
``local flavor unification''~\cite{Baur:2019iai}. This holistic picture teaches us some general lessons:
\begin{itemize}
\item One cannot just consider a specific flavor group (e.g.\ modular flavor) without the others.

\item There is always a traditional flavor group (universal in moduli space) that could give 
severe restrictions to the K\"ahler potential and superpotential of the theory.

\item Apart from the finite discrete modular flavor group and its specific representations, 
one has to consider the modular weights of the matter fields as well, as they might lead to 
further $R$-symmetries that play the role of ``shaping symmetries''.
\end{itemize}

Flavor symmetries have to be spontaneously broken and thus the full eclectic picture requires 
several sources of breakdown. On the one hand this leads to a serious complication of the picture. 
On the other hand, it is welcome once we want to obtain the hierarchical structure of masses and 
mixing angles of quarks \emph{and} leptons. While the modular group can be broken via the 
moduli (with hierarchical patterns close to the fixed points of $\SL{2,\Z{}}$), we have to 
consider additional flavon fields to break the traditional flavor symmetries via nontrivial vacuum 
expectation values (vevs). These flavon vevs might then lead to a further breakdown of the discrete 
modular flavor symmetry as well. It is this subtle interplay of breakdown via moduli and flavons 
that is the main subject of the present paper. We shall see that there are various ways to obtain 
hierarchical patterns from the breakdown of the eclectic flavor group via moduli and flavons.

Flavor symmetries should, of course, also be discussed from a BU perspective. 
There one has the free choice of the groups, the representations and modular weights of matter 
fields to confront a model with existing data. Ideally one might hope to find a specific model as a 
``best fit'' to masses and mixing angles of quarks and leptons. This then might give useful hints 
towards a fundamental theory of flavor. Unfortunately, no one has yet been able to identify such a 
distinct model (or even a class of models). Good fits to the data can be achieved for various 
groups and representations. Still, even in the BU approach, we might try to find some theoretical 
guidelines. If we, for example, have a traditional flavor group $G$, we could design an eclectic 
scheme with a discrete modular flavor group that is included in the outer automorphisms of the 
group $G$~\cite{Nilles:2020nnc}. This could be a first step to bridge the gap between the TD and 
BU approaches to flavor. Both approaches enjoy various desirable properties as e.g.\ the 
appearance of local flavor unification with enhanced symmetry at certain fixed points or regions 
in moduli space, eventually broken spontaneously via moduli vevs.

The TD approach is very restrictive and a challenge for realistic model building:
\begin{itemize}
\item We first have to design models with the desired flavor groups explicitly in 
string compactifications.

\item The explicit representations of the flavor group are then fixed and cannot 
be chosen by hand.

\item Likewise, modular weights are fixed and determined as well.

\item There are various restrictions from $R$-symmetries that appear in the 
six-dimensional compactification.
\end{itemize}
Given these restrictions, there remains still a wide gap between TD attempts and the models 
considered in the BU approach.

There have not been any attempts yet in TD model building, and
in this paper, we want to make a first step in this direction.
We are particularly interested in string models with elliptic
fibrations, and these are classified according to two-dimensional
$\mathbbm{T}^2/\Z{k}$ orbifold sectors with $k=2,3,4,6$. The traditional flavor
symmetries of these scenarios were analyzed some time ago~\cite{Kobayashi:2006wq,Olguin-Trejo:2018wpw}. 
In those works, it is found that the $\mathbbm{T}^2/\Z3$ orbifold sector 
(which in some sense also qualitatively covers the $\Z6$ case as well)
leads to the most promising class of string models that reproduce the matter
spectrum of the minimal supersymmetric standard model (MSSM). These models are endowed 
with traditional flavor symmetry $\Delta(54)$ and twisted states
that transform as (irreducible) triplet representations of
this group. Therefore, we want to concentrate on this class
of models with traditional flavor symmetry $\Delta(54)$ and
modular flavor symmetry $T'$. Fortunately, there have
been explicit semirealistic model constructions of heterotic string theory
that exhibit elliptic fibrations of type $\Z3$~\cite{Carballo-Perez:2016ooy}. 
A full classification of these models with the relevant field content
is given in table~\ref{tab:Z3xZ3configurations} of the present paper.

All of these models share the same eclectic flavor group but
differ in the available representations of candidate
flavon and matter fields. A first step toward phenomenological
applications would then be the analysis of breakdown patterns of the
eclectic flavor group and this is the main goal of the
present paper. We have to clarify which fields are needed
for an efficient breakdown of the eclectic flavor group which,
in addition, allow a hierarchical pattern for a successful
description of masses and mixing angles of quarks and
leptons (originating partially through the proximity to
local gauge group enhancements). Before proceeding to
explicit model building, we would therefore like to know the qualitative
breakdown pattern of $\Delta(54)$ and $T'$. In future
work~\cite{Baur:2021pr} we shall then use these results for the selection
of suitable models (from the classes  
in table~\ref{tab:Z3xZ3configurations}) for phenomenological applications.
 
The paper is structured as follows:
In section~\ref{sec:T2overZ3}, we describe a $\mathbbm{T}^2/\Z{3}$ orbifold sector with traditional 
flavor group $\Delta(54)$ and eclectic flavor group\footnote{We follow the notation of the SmallGroup 
library of GAP~\cite{GAP4}, where the first number in the square parentheses denotes the 
order of the group and the second number is the group id. We also use the nomenclature conventions 
of~\cite{Jurciukonis:2017mjp}.} $\Omega(2) \cong [1944,3448]$ (as the multiplicative closure of 
$\Delta(54)$, $T^\prime$ and $\Z{9}^R$), derived as the two-dimensional elliptic fibration 
of a $\mathbbm{T}^6/(\Z{3}\x\Z{3})$ orbifold. We identify the representations of the groups 
in the massless sector including the modular weights. At a generic point in moduli space, we have 
a flavor symmetry $\Delta(54)\cup\Z{9}^R =\Delta'(54,2,1)\cong[162,44]$. We discuss in detail 
the enhancements at the fixed points of the K\"ahler modulus $T=\I$, $1$, and $\omega:=\exp\left(\nicefrac{2\pi \I}{3}\right)$ 
that lead to the groups $\Xi(2,2)\cong[324,111]$ (for $T=\I$) or [468,125] (for $T=1, \omega$). 
In section~\ref{sec:Delta54Breaking} 
we discuss the possible breakdown of $\Delta(54)$ via flavon vevs. Candidate flavons are identified 
via the inspection of the massless sector of the $\mathbbm{T}^6/(\Z{3}\times\Z{3})$ orbifold under 
consideration. We show that triplets of $\Delta(54)$ are very efficient in the breakdown of the traditional 
flavor group. We extend this analysis to the breakdown of the eclectic flavor group in section~\ref{sec:Omega2Breaking}, 
with particular attention to the groups [324,111] and [468,125] at the fixed points $T=\I$ and 
$T=1,\omega$, respectively. The various breakdown patterns are summarized in the 
figures~\ref{fig:D54breaking},~\ref{fig:Xi22breaking}, and~\ref{fig:H321breaking} 
with details given 
in tables~\ref{tab:D54breaking},~\ref{tab:Xi22}, and \ref{tab:H321}. 
In section~\ref{sec:conclusions} we conclude by discussing the relevance of our analysis to 
flavor model building and give an outlook to future work.

\section{Holistic picture of the eclectic flavor symmetry}
\label{sec:T2overZ3}
    
Eclectic flavor groups~\cite{Nilles:2020nnc} arise naturally in  compactifications of 
string theory~\cite{Nilles:2020kgo,Nilles:2020tdp,Baur:2021mtl,Ohki:2020bpo}. They consist 
of a nontrivial combination of the effective traditional flavor symmetries $G_\mathrm{traditional}$
and finite modular symmetries $G_\mathrm{modular}$ under which string states are charged.
Eclectic flavor groups can also include \CP-like transformations and discrete $R$-symmetries. 
Let us here discuss the appearance of these symmetries in detail.

We focus on factorizable six-dimensional orbifolds, which contain
three two-dimensional $\mathbbm T^2/\Z{N}$ orbifold sectors. Each of the $\mathbbm T^2$
is endowed with two modular groups, $\SL{2,\Z{}}_U$ and $\SL{2,\Z{}}_T$, associated 
respectively with the  complex structure modulus $U$ and the (stringy) K\"ahler modulus $T$
of the torus. For $N=2$, the values of both $T$ and $U$ remain unrestricted.
For $N>2$, due to its geometric nature, $U$ has to be  fixed to a value that is compatible
with the orbifold twist \Z{N}. 
In these cases $\SL{2,\Z{}}_U$ is broken down 
to a discrete remnant of the Lorentz group of discrete rotations in the compact dimensions, which 
is an Abelian group compatible with the orbifold. This subgroup appears as a discrete 
$R$-symmetry in the effective theory~\cite{Nilles:2020tdp}. 
$\SL{2,\Z{}}_T$, however,  remains a symmetry of the effective theory. This symmetry is nonlinearly 
realized, as can be seen from its action on the modulus $T$ and matter fields $\Phi_{n}$.
An element $\gamma\in\SL{2,\Z{}}_T$ transforms these fields according to~\cite{Lauer:1989ax,Lauer:1990tm}
\begin{equation}
\label{eq:modularTrafo}
 T ~\xmapsto{\gamma}~ \frac{aT+b}{cT+d}\quad\text{and}\quad
 \Phi_n ~\xmapsto{\gamma}~ (cT+d)^n \rho_{\rep s}(\gamma) \Phi_n\,,\quad\text{with}\quad
 \gamma:=\begin{pmatrix}a&b\\c&d\end{pmatrix}\in\SL{2,\Z{}}_T\,.
\end{equation}
Here, $\Phi_n$ denotes a multiplet of string matter states with identical quantum numbers,
except for their location at different orbifold singularities.
Furthermore, $n$ denotes the modular weight of the matter multiplets $\Phi_n$, 
$(cT+d)^n$ is known as automorphy factor, and $\rho_{\rep s}(\gamma)$ is an $s$-dimensional 
matrix representation of $\gamma$ in a (discrete) finite modular group $G_\mathrm{modular}$
that depends on the \Z{N} twist of the orbifold sector.

On the other hand, the traditional flavor symmetries $G_\mathrm{traditional}$ can be identified 
through the geometric features of toroidal orbifolds~\cite{Kobayashi:2006wq}. First, 
$G_\mathrm{traditional}$ includes the permutations among the various equivalent orbifold 
singularities where matter states comprising the multiplets $\Phi_n$ are located. Since
permutations are non-Abelian, so are these traditional flavor symmetries. Second, 
$G_\mathrm{traditional}$ contains the discrete symmetries governing the admissible 
couplings among the states in $\Phi_n$, which are known as string selection rules. The resulting
traditional flavor group is obtained by  multiplicative closure of these two types of
symmetries. It is universal in moduli space, as all its elements only act on $\Phi_n$.

In addition to traditional flavor and classical modular transformations, there are \CP-like transformations. 
On the moduli $T$ and $U$ these act with an element $\gamma$ of determinant $-1$, thereby enhancing the 
respective modular groups to $\SL{2,\Z{}}_{T,U}\rtimes\Z{2}\cong\GL{2,\Z{}}_{T,U}$. Fixing $U$ by 
orbifolding selects specific \CP-like transformations compatible with the twist \Z{N} and 
fixed value of $U$~\cite{Baur:2019iai,Nilles:2020gvu}.
The general action on the modulus $T$ is given by~\cite{Baur:2019kwi,Baur:2019iai} 
(see also~\cite{Nilles:2020gvu} as well as~\cite{Novichkov:2019sqv} for the BU approach)
\begin{equation}
\label{eq:CPTrafoMod}
 T ~\xmapsto{\gamma_{\CP}}~ \frac{a\bar{T}+b}{c\bar{T}+d}\,,\quad\text{with}\quad
 \gamma_{\CP}:=\begin{pmatrix}a&b\\c&d\end{pmatrix}\in\GL{2,\Z{}}_T\quad\text{and}\quad \det \gamma_\CP=-1,
\end{equation}
while string matter multiplets transform as
\begin{equation}
\label{eq:CPTrafoFields}
 \Phi_n ~\xmapsto{\gamma_{\CP}}~ (c\bar{T}+d)^n \rho_{\rep{\bar{s}}}(\gamma_{\CP}) \bar{\Phi}_n\;.
\end{equation}
Here, bars denote complex conjugation.\footnote{We note that, in general, not all of the transformations 
$\gamma_\CP$ correspond to transformations that map representations of string matter multiplets 
$\rep{s}$ to their complex conjugate representations $\rep{\bar{s}}$. Whether or not this is the 
case crucially depends on the nature of the outer automorphisms of the involved finite groups 
and the specific representations~\cite{Chen:2014tpa}. However, as we will consider throughout 
this work only the \textit{massless} spectrum of $\mathbbm T^2/\Z3$, for which complex conjugation 
applies~\cite{Nilles:2018wex}, we restrict ourselves to the statement of eq.~\eqref{eq:CPTrafoFields}.}

Interestingly, with the help of the Narain formulation of toroidal orbifolds~\cite{Narain:1985jj,Narain:1986am}
(see~\cite{GrootNibbelink:2017usl} for further technical details),
it is found that the outer automorphisms of the orbifold space group yield all symmetries
of the effective theory of these compactifications. In particular, all of $G_\mathrm{traditional}$ 
and $G_\mathrm{modular}$, as well as the discrete $R$-symmetries and the \CP-like transformations 
turn out to have their origin in these outer automorphisms, revealing a unified origin of flavor in string 
compactifications~\cite{Baur:2019kwi}.

This rich set of flavor symmetries containing all these elements builds the
eclectic flavor group of a toroidal orbifold. Thus, an eclectic flavor group is the 
multiplicative closure of the various flavor subgroups, i.e.
\begin{equation}
\label{eq:generalEclectic}
 G_\mathrm{eclectic} ~=~ G_\mathrm{traditional} ~\cup~ G_\mathrm{modular} ~\cup~ G_\mathrm{R} ~\cup~ \CP\,,
\end{equation}
where $G_\mathrm{R}$ denotes here the Abelian discrete $R$-symmetry, remnant of $\SL{2,\Z{}}_U$.

One of the phenomenological advantages of models endowed with eclectic flavor
symmetries is that they provide control over the structure of the effective
superpotential and K\"ahler potential, preventing in particular the loss of
predictability that is found in models based on finite modular symmetries 
only~\cite{Chen:2019ewa}. The predictive power of the eclectic picture results from 
the large amount of symmetry of $G_\mathrm{eclectic}$. 
Phenomenological applications require a spontaneous breakdown of this huge
symmetry, and this is a challenge for flavor model building.
One of the goals of this paper is precisely to address this question
in an illustrative example of the eclectic picture based on string theory.

\subsection[The eclectic flavor symmetry of T2/Z3]{\boldmath The eclectic flavor symmetry of $\mathbbm T^2/\Z3$ \unboldmath}

We chose the  $\mathbbm T^2/\Z3$ orbifold sector as an example to illustrate the phenomenological
potential of eclectic flavor symmetries. The outer automorphisms of the corresponding Narain 
space group yield the following symmetries~\cite{Baur:2019iai,Nilles:2020kgo,Nilles:2020tdp}:
\begin{itemize}
\item an $\SL{2,\Z{}}_T$ modular symmetry which acts as a $\Gamma'_{3}\cong T'$ finite 
      modular symmetry on matter fields and their couplings,
\item a $\Delta(54)$ traditional flavor symmetry,
\item a $\Z9^R$ discrete $R$-symmetry as remnant of $\SL{2,\Z{}}_U$, and 
\item a $\Z2^\CP$ \CP-like transformation.
\end{itemize}
As explained in detail in~\cite{Nilles:2020tdp,Nilles:2020gvu} and summarized in 
table~\ref{tab:Z3FlavorGroups}, the $\SL{2,\Z{}}_T$ modular generators $\mathrm{S}$ and 
$\mathrm{T}$ arise from rotational outer automorphisms. These generators can be 
represented by
\begin{equation}
  \mathrm S ~=~ \begin{pmatrix} 0 & 1 \\ -1 & 0 \end{pmatrix}\qquad\text{and}\qquad 
  \mathrm T ~=~ \begin{pmatrix} 1 & 1 \\  0 & 1 \end{pmatrix}\,,
\end{equation}
and act on the modulus $T$ and the matter fields $\Phi_n$ according to eq.~\eqref{eq:modularTrafo}.
Further, there is a reflectional outer automorphism which corresponds to a $\Z2^\CP$ \CP-like 
transformation. It can be chosen to be represented by
\begin{equation}\label{eq:Kstar}
 \mathrm K_* ~=~ \begin{pmatrix} -1 & 0 \\ 0 & 1 \end{pmatrix}\;,
\end{equation}
which acts on the modulus and matter fields as in eq.~\eqref{eq:CPTrafoMod}.

The traditional flavor symmetry $\Delta(54)$ is generated by two translational outer automorphisms 
of the Narain space group $\mathrm A$ and $\mathrm B$ of order 3 together with the \Z2 rotational 
outer automorphism $\mathrm C := \mathrm S^2$.
The automatic identification of $\mathrm C$ with $\mathrm S^2$ implies that $\Delta(54)$ and the 
modular symmetry have a nontrivial overlap and, furthermore, that $\Delta(54)$ is actually a 
non-Abelian $R$-symmetry.

Finally, in our example, the complex structure modulus is geometrically 
stabilized at $\vev{U}=\exp(\nicefrac{2\pi\I}3)$ in order for $\mathbbm T^2$ to be 
compatible with the \Z3 point group. If $\mathbbm T^2/\Z3$ is embedded in a 
six-dimensional orbifold, $\vev{U}$ breaks the original $\SL{2,\Z{}}_U$ of $\mathbbm T^2$
to a discrete remnant generated by $\mathrm R$ that acts as a $\Z9^R$ symmetry on matter 
fields (normalizing their $R$-charges to be integers). 

Since $\Delta(54)$ and $\Z9^R$ both act trivially on the modulus $T$,  
the traditional flavor symmetry is enhanced to $\Delta(54)\cup\Z9^R\cong\Delta'(54,2,1)\cong [162,44]$. 
Including the additional $\Z2^\CP$ \CP-like transformation~\eqref{eq:CPTrafoFields} the full eclectic group 
according to eq.~\eqref{eq:generalEclectic} is a group of order $3888$ given by
\begin{equation}
  G_\mathrm{eclectic} ~=~ \Omega(2)\rtimes\Z2^\CP\,,\qquad\text{where}\quad \Omega(2)\cong [1944,3448]\,.
\end{equation}

\begin{table}[t!]
\center
\resizebox{\textwidth}{!}{
\begin{tabular}{|c|c||c|c|c|c|c|c|}
\hline
\multicolumn{2}{|c||}{nature}        & outer automorphism       & \multicolumn{5}{c|}{\multirow{2}{*}{flavor groups}} \\
\multicolumn{2}{|c||}{of symmetry}   & of Narain space group    & \multicolumn{5}{c|}{}\\
\hline
\hline
\parbox[t]{3mm}{\multirow{6}{*}{\rotatebox[origin=c]{90}{eclectic}}} &\multirow{2}{*}{modular}            & rotation $\mathrm{S}~\in~\SL{2,\Z{}}_T$ & $\Z{4}$      & \multicolumn{3}{c|}{\multirow{2}{*}{$T'$}} &\multirow{6}{*}{$\Omega(2)$}\\
&                                    & rotation $\mathrm{T}~\in~\SL{2,\Z{}}_T$ & $\Z{3}$      & \multicolumn{3}{c|}{}                      & \\
\cline{2-7}
&                                    & translation $\mathrm{A}$                & $\Z{3}$      & \multirow{2}{*}{$\Delta(27)$} & \multirow{3}{*}{$\Delta(54)$} & \multirow{4}{*}{$\Delta'(54,2,1)$} & \\
& traditional                        & translation $\mathrm{B}$                & $\Z{3}$      &                               & & & \\
\cline{3-5}
& flavor                             & rotation $\mathrm{C}=\mathrm{S}^2\in\SL{2,\Z{}}_T$      & \multicolumn{2}{c|}{$\Z{2}^R$} & & & \\
\cline{3-6}
&                                    & rotation $\mathrm{R}\in\SL{2,\Z{}}_U$   & \multicolumn{3}{c|}{$\Z{9}^R$}   & & \\
\hline
\end{tabular}
}
\caption{\label{tab:Z3FlavorGroups}
Eclectic flavor group $\Omega(2)$ for six-dimensional orbifolds that contain a 
$\mathbbm T^2/\Z{3}$ orbifold sector. In this case, $\SL{2,\Z{}}_U$ of the stabilized complex 
structure modulus $\vev U=\exp\left(\nicefrac{2\pi\I}{3}\right)$ is broken, resulting in a remnant 
$\Z{9}^R$ $R$-symmetry. Including $\Z{9}^R$ enhances the traditional flavor group $\Delta(54)$ to 
$\Delta'(54,2,1)\cong [162,44]$, which together with $T'$ finally leads to the eclectic group 
$\Omega(2) \cong [1944, 3448]$. Table from~\cite{Nilles:2020tdp}.}
\end{table}

In semirealistic heterotic orbifold compactifications endowed with a $\mathbbm T^2/\Z3$ 
orbifold sector, the massless spectrum consists of (i) untwisted string states that transform
as flavor singlet states $\Phi_0$ or $\Phi_{-1}$ free to move in the bulk, and (ii) 
twisted string states transforming as flavor triplets, attached to the three orbifold 
fixed points. The modular weights of twisted states depend on the twisted sector they 
belong to. As shown in table~\ref{tab:Representations}, in the $\theta$ sector
only $n\in\{\nicefrac{-2}{3},\nicefrac{-5}{3}\}$ are possible, while only 
$n\in\{\nicefrac{-1}{3},\nicefrac23\}$ appear in the $\theta^2$ twisted sector.
The transformations~\eqref{eq:modularTrafo} of twisted triplets $\Phi_{\nicefrac{-2}{3}}$
are governed by the three-dimensional matrix representations of the modular 
generators~\cite{Baur:2019iai}
\begin{equation}
\label{eq:ModularRepresentation}
\rho(\mathrm S) = \frac{\I}{\sqrt3}\begin{pmatrix} 1&1&1\\1&\omega^2&\omega\\1&\omega&\omega^2\end{pmatrix}
\qquad\text{and}\qquad
\rho(\mathrm T) = \begin{pmatrix} \omega^2&0&0\\0&1&0\\0&0&1\end{pmatrix}\,,
\end{equation}
which form the representation $\rep s=\rep2'\oplus\rep1$ of the finite modular group 
$\Gamma_3'\cong T'\cong[24,3]$. Furthermore, the action of the $\Delta(54)$ generators on $\Phi_{\nicefrac{-2}{3}}$ is given by
\begin{equation}
\label{eq:TraditionalRepresentation}
\rho(\mathrm A) = \begin{pmatrix} 0&1&0\\0&0&1\\1&0&0\end{pmatrix}\,,~~
\rho(\mathrm B) = \begin{pmatrix} 1&0&0\\0&\omega&0\\0&0&\omega^2\end{pmatrix}
~~\text{and }~~
\rho(\mathrm C) = -\begin{pmatrix} 1&0&0\\0&0&1\\0&1&0\end{pmatrix} ~=~  \rho(\mathrm S)^2 \,,
\end{equation}
and they generate the representation $\rep r = \rep 3_2$ of $\Delta(54)$.
The representations associated with the other twisted triplets are expressed in terms
of eqs.~\eqref{eq:ModularRepresentation} and~\eqref{eq:TraditionalRepresentation},
according to the prescription given in table~\ref{tab:Representations}. 
Finally, the integer $\Z9^R$ $R$-charges of matter fields can be uniquely determined 
by using their $\SL{2,\Z{}}_U$ properties and the fixed value $\vev{U}$, 
see~\cite[sec.~4.2]{Nilles:2020gvu}.

It is important to stress that the massless spectrum of the $\mathbb T^2/\Z3$ orbifold
sector of heterotic orbifold compactifications does not include all possible representations
of $\Delta(54)$. In particular, note that there are no massless doublet representations. 
In fact, doublets arise as winding modes of strings around different singularities and, 
hence, correspond to massive states~\cite{Nilles:2018wex}.
    
\begin{table}[t!]
\center
\resizebox{\textwidth}{!}{
\begin{tabular}{|c||c||c|c|c|c||c|c|c|c||c|}
\hline
\multirow{3}{*}{sector} &\!\!matter\!\!& \multicolumn{9}{c|}{eclectic flavor group $\Omega(2)$}\\
                        &fields        & \multicolumn{4}{c||}{modular $T'$ subgroup} & \multicolumn{4}{c||}{traditional $\Delta(54)$ subgroup} & $\Z{9}^R$ \\
                        &$\Phi_n$      & \!\!irrep $\rep{s}$\!\! & $\rho_{\rep{s}}(\mathrm{S})$ & $\rho_{\rep{s}}(\mathrm{T})$ & $n$ & \!\!irrep $\rep{r}$\!\! & $\rho_{\rep{r}}(\mathrm{A})$ & $\rho_{\rep{r}}(\mathrm{B})$ & $\rho_{\rep{r}}(\mathrm{C})$ & $R$\\
\hline
\hline
bulk      & $\Phi_{\text{\tiny 0}}$   & $\rep1$             & $1$                   & $1$                   & $0$               & $\rep1$   & $1$               & $1$                   & $+1$ & $0$         \\
          & $\Phi_{\text{\tiny $-1$}}$& $\rep1$             & $1$                   & $1$                   & $-1$              & $\rep1'$  & $1$               & $1$                   & $-1$ & $3$         \\
\hline
$\theta$  & $\Phi_{\nicefrac{-2}{3}}$ & $\rep2'\oplus\rep1$ & $\rho(\mathrm{S})$    & $\rho(\mathrm{T})$    & $\nicefrac{-2}{3}$& $\rep3_2$ & $\rho(\mathrm{A})$& $\rho(\mathrm{B})$    & $+\rho(\mathrm{C})$ & $1$\\
          & $\Phi_{\nicefrac{-5}{3}}$ & $\rep2'\oplus\rep1$ & $\rho(\mathrm{S})$    & $\rho(\mathrm{T})$    & $\nicefrac{-5}{3}$& $\rep3_1$ & $\rho(\mathrm{A})$& $\rho(\mathrm{B})$    & $-\rho(\mathrm{C})$ & $-2$\\
\hline
$\theta^2$& $\Phi_{\nicefrac{-1}{3}}$ & $\rep2''\oplus\rep1$& $(\rho(\mathrm{S}))^*$& $(\rho(\mathrm{T}))^*$& $\nicefrac{-1}{3}$& $\crep3_1$& $\rho(\mathrm{A})$& $(\rho(\mathrm{B}))^*$& $-\rho(\mathrm{C})$ & $2$\\
          & $\Phi_{\nicefrac{+2}{3}}$ & $\rep2''\oplus\rep1$& $(\rho(\mathrm{S}))^*$& $(\rho(\mathrm{T}))^*$& $\nicefrac{+2}{3}$& $\crep3_2$& $\rho(\mathrm{A})$& $(\rho(\mathrm{B}))^*$& $+\rho(\mathrm{C})$ & $5$\\
\hline
\hline
super-    & \multirow{2}{*}{$\mathcal{W}$} & \multirow{2}{*}{$\rep1$} & \multirow{2}{*}{$1$} & \multirow{2}{*}{$1$} & \multirow{2}{*}{$-1$} & \multirow{2}{*}{$\rep1'$} & \multirow{2}{*}{$1$} & \multirow{2}{*}{$1$} & \multirow{2}{*}{$-1$} & \multirow{2}{*}{$3$}\\
\!\!potential\!\! & & & & & & & & & & \\
\hline
\end{tabular}
}
\caption{\label{tab:Representations}
$T'$ and $\Delta(54)$ irreducible representations and $\Z9^R$ $R$-charges of massless
matter fields $\Phi_n$ with modular weights $n$ in semirealistic heterotic orbifold compactifications 
with a $\mathbbm{T}^2/\Z3$ sector. $T'$, $\Delta(54)$ and $\Z9^R$ combine nontrivially to the 
eclectic flavor group $\Omega(2) \cong [1944, 3448]$, as described in table~\ref{tab:Z3FlavorGroups}. 
Untwisted matter fields $\Phi_n$ (with integer modular weights $n$) form one-dimensional 
representations, while twisted matter fields $\Phi_n$ (with fractional modular weights $n$) 
build triplet representations. Table from \cite{Nilles:2020kgo}.}
\end{table}
    
At different points $\vev{T}$ in moduli space, $\SL{2,\Z{}}_T$ and $T'$ are broken down
to the stabilizer subgroup, i.e.\ to the modular subgroup that leaves $\vev T$ invariant. Since
the modulus is no longer transformed at $\vev T$, the surviving symmetry yields an enhancement 
of the traditional flavor symmetry $\Delta(54)$. This enhancement, associated with the specific 
location in moduli space, has been called {\it local flavor unification}~\cite{Baur:2019iai}. 
Next, we study the most relevant scenarios of local flavor unification in the  $\mathbbm T^2/\Z3$ 
orbifold sector.

\subsection[Flavor enhancement at T=i]{\boldmath Flavor enhancement at $\vev T = \I$ \unboldmath}
\label{sec:FlavorEnhancementT=i}

First, we discuss the point $\langle T\rangle=\I$ in moduli space (see section 6.2 in ref.~\cite{Nilles:2020gvu} 
for details of the derivation). In this case, $\SL{2,\Z{}}_T$ is broken to a $\Z{4}$ subgroup generated 
by $\mathrm{S}$. Twisted matter fields transform as
\begin{subequations}\label{eq:ModularS}
\begin{eqnarray}
\Phi_{\nicefrac{-2}{3}} & \xmapsto{~\mathrm{S}~} & \exp(\nicefrac{2\pi\I}{6})\, \rho(\mathrm{S})\,\Phi_{\nicefrac{-2}{3}}\label{eq:ModularS32}\;,\\
\Phi_{\nicefrac{-5}{3}} & \xmapsto{~\mathrm{S}~} & \exp(\nicefrac{2\pi\I\, 5}{12})\, \rho(\mathrm{S})\,\Phi_{\nicefrac{-5}{3}}\label{eq:ModularS31}\;.
\end{eqnarray}
\end{subequations}
Here we use $c=-1$, $d=0$ for the automorphy factor $(c\langle T\rangle+d)^{n}=(-\I)^{n}$ of 
$\mathrm{S}\in\SL{2,\Z{}}_T$ at $\langle T\rangle=\I$, which results in
\begin{subequations}\label{eq:automorphyI}
\begin{align}
 (-\I)^{\nicefrac{-2}{3}} &= \exp(\nicefrac{2\pi\I}{6})&      &\text{for } n ~=~ \nicefrac{-2}{3}\;,&\\
 (-\I)^{\nicefrac{-5}{3}} &= \exp(\nicefrac{2\pi\I\, 5}{12})& &\text{for } n ~=~ \nicefrac{-5}{3}\;.&
\end{align}
\end{subequations}
The superpotential also transforms under $\mathrm{S}$ as 
$\mathcal{W} \xmapsto{\mathrm{S}} \I\,\mathcal{W}$. This implies that also $\mathrm{S}$ generates 
a discrete $R$-symmetry, which altogether leads to a non-Abelian discrete $R$-symmetry~\cite{Chen:2013dpa}.

At $\langle T\rangle=\I$, the explicit representation matrices (which include the associated automorphy factors) 
of the unified flavor group of twisted matter fields $\Phi_{\nicefrac{-2}{3}}$ and $\Phi_{\nicefrac{-5}{3}}$ are given by
\begin{align}
 &\Phi_{\nicefrac{-2}{3}}:& 
 \rho_{\rep{3}_2,\I}(\mathrm{A})&=\rho(\mathrm A),&
 \rho_{\rep{3}_2,\I}(\mathrm{B})&=\rho(\mathrm B),&
 \rho_{\rep{3}_2,\I}(\mathrm{C})&=\rho(\mathrm C),& \\\notag
 && 
 \rho_{\rep{3}_2,\I}(\mathrm{R})&=\e^{\nicefrac{2\pi\I}{9}}\,\Id_3,&
 \rho_{\rep{3}_2,\I}(\mathrm{S})&=\e^{\nicefrac{2\pi\I}{6}}\,\rho(\mathrm S),&\text{and}\\
 &\Phi_{\nicefrac{-5}{3}}:&  
 \rho_{\rep{3}_1,\I}(\mathrm{A})&=\rho(\mathrm A),&
 \rho_{\rep{3}_1,\I}(\mathrm{B})&=\rho(\mathrm B),&
 \rho_{\rep{3}_1,\I}(\mathrm{C})&=-\rho(\mathrm C),& \\
 \notag
 &&
 \rho_{\rep{3}_1,\I}(\mathrm{R})&=\e^{\nicefrac{-4\pi\I}{9}}\,\Id_3,&
 \rho_{\rep{3}_1,\I}(\mathrm{S})&=\e^{\nicefrac{2\pi\I\, 5}{12}}\,\rho(\mathrm S).&
\end{align}
The $\CP$-like transformation generated by eq.~\eqref{eq:Kstar} acts on the modulus as 
$T\mapsto-\bar{T}$, which is conserved for $\langle T\rangle=\I$. The corresponding 
automorphy factor and representation matrix in eqs.~\eqref{eq:CPTrafoMod} and~\eqref{eq:CPTrafoFields}
are trivial, see~\cite{Baur:2019kwi,Nilles:2020gvu}, such that\footnote{For explicit computations, 
it is useful to combine $\Phi\oplus\bar{\Phi}$ and extend also the other generators to this
$\rep{r}\oplus\rep{\bar{r}}$-dimensional representation.} $\Phi_n\xmapsto{\CP}\bar{\Phi}_n$.

Altogether, the linearly realized unified flavor group at $\langle T\rangle=\I$ is found to be 
\begin{equation}
\label{eq:groupTi}
\Delta(54) ~\cup~ \Z{9}^R ~\cup~ \mathrm{S} ~\cup~ \Z2^\CP~=~ \Xi(2,2)\rtimes\Z2^\CP ~\cong~ [324,111]\rtimes\Z2~\cong~ [648,548]\;.
\end{equation}

\subsection[Flavor enhancement at T=omega]{\boldmath Flavor enhancement at $\vev T = \omega$ \unboldmath}
\label{sec:FlavorEnhancementT=w}

Next, following section 6.3 in ref.~\cite{Nilles:2020gvu}, we consider the point $\langle T\rangle=\omega$ in moduli 
space. There, $\SL{2,\Z{}}_T$ is broken to a $\Z{3}$ subgroup generated by $\mathrm{S}\mathrm{T}$ 
such that our twisted matter fields transform as
\begin{subequations}\label{eq:ModularST}
\begin{eqnarray}
\Phi_{\nicefrac{-2}{3}} & \xmapsto{~\mathrm{ST}~} & \exp(\nicefrac{2\pi\I\, 2}{9})\, \rho(\mathrm{S}\mathrm{T})\,\Phi_{\nicefrac{-2}{3}}\label{eq:ModularST32}\;,\\
\Phi_{\nicefrac{-5}{3}} & \xmapsto{~\mathrm{ST}~} & \exp(\nicefrac{2\pi\I\, 5}{9})\, \rho(\mathrm{S}\mathrm{T})\,\Phi_{\nicefrac{-5}{3}}\label{eq:ModularST31}\;.
\end{eqnarray}
\end{subequations}
Here we use $c=d=-1$ for $\mathrm{S}\mathrm{T}\in\SL{2,\Z{}}_T$ which yields 
$(c\langle T\rangle+d)^{n}=(-\omega-1)^{n}=(\omega^2)^{n}$ and, hence, the automorphy 
factors
\begin{subequations}\label{eq:automorphyOmega}
\begin{align}
(\omega^2)^{\nicefrac{-2}{3}} &= \exp(\nicefrac{2\pi\I\, 2}{9})& &\mathrm{for}\ n ~=~ \nicefrac{-2}{3}\;,&\\
(\omega^2)^{\nicefrac{-5}{3}} &= \exp(\nicefrac{2\pi\I\, 5}{9})& &\mathrm{for}\ n ~=~ \nicefrac{-5}{3}\;.&
\end{align}
\end{subequations}
In addition, the superpotential transforms under $\mathrm{S}\mathrm{T}$ as 
$\mathcal{W} \xmapsto{\mathrm{S}\mathrm{T}} \omega\,\mathcal{W}$. 
Hence, also here the residual flavor symmetry is a non-Abelian 
discrete $R$-symmetry~\cite{Chen:2013dpa}.

The explicit representation matrices of the unified flavor group of twisted matter fields 
$\Phi_{\nicefrac{-2}{3}}$ and $\Phi_{\nicefrac{-5}{3}}$ are given by
\begin{align}
 \label{eq:GenTwPhi23}
 &\Phi_{\nicefrac{-2}{3}}:& 
 \rho_{\rep{3}_2,\omega}(\mathrm{A})&=\rho(\mathrm A),&
 \rho_{\rep{3}_2,\omega}(\mathrm{B})&=\rho(\mathrm B),&
 \rho_{\rep{3}_2,\omega}(\mathrm{C})&=\rho(\mathrm C),& \\
 \notag
 && 
 \rho_{\rep{3}_2,\omega}(\mathrm{R})&=\e^{\nicefrac{2\pi\I}{9}}\,\Id_3,&
 \rho_{\rep{3}_2,\omega}(\mathrm{ST})&=\e^{\nicefrac{2\pi\I\,2}{9}}\,\rho(\mathrm{ST}), &\text{and}  \\
 \label{eq:GenTwPhi53}
 &\Phi_{\nicefrac{-5}{3}}:&  
 \rho_{\rep{3}_1,\omega}(\mathrm{A})&=\rho(\mathrm A),&
 \rho_{\rep{3}_1,\omega}(\mathrm{B})&=\rho(\mathrm B),&
 \rho_{\rep{3}_1,\omega}(\mathrm{C})&=-\rho(\mathrm C),& \\\notag
 &&
 \rho_{\rep{3}_1,\omega}(\mathrm{R})&=\e^{\nicefrac{-4\pi\I}{9}}\,\Id_3,&
 \rho_{\rep{3}_1,\omega}(\mathrm{ST})&=\e^{\nicefrac{2\pi\I\,5}{9}}\,\rho(\mathrm{ST}).&
\end{align}
A representative of the $\Z2^\CP$ transformation is given by $\mathrm{K}_*\mathrm T$,
which acts on the modulus and matter fields as 
\begin{align}\label{eq:CPtrafoTomega}
&T\xmapsto{\CP}-\bar{T}-1\;,& &\text{and}& &\Phi_n\xmapsto{\CP}\rho(\mathrm{T})^*\bar{\Phi}_n\;,&
\end{align}
such that $\langle T\rangle=\omega$ is left invariant.

The linearly realized unified flavor group at $\vev T =\omega$ is thus
\begin{equation}
\label{eq:UnifiedFlavorGroupT=w}
\Delta(54) ~\cup~ \Z{9}^R ~\cup~ \mathrm{S}\mathrm{T} ~\cup~ \Z2^\CP 
~=~ H(3,2,1)\rtimes\Z2^\CP ~\cong~ [486,125]\rtimes\Z2 ~\cong~ [972,469]\;.
\end{equation}

\subsection[Flavor enhancement at T=1 and T=I*infinity]{\boldmath Flavor enhancement at $\vev T = 1$ and $\vev T = \I\,\infty$\unboldmath}
\label{sec:FlavorEnhancementT=1}

Since the point $\vev T = 1$ has not been investigated in the literature for residual symmetries,
we give more details here. First, note that the discussion for $\vev T = 1$ also applies to the point 
$\vev T = \I\,\infty$ because these points are dual via conjugation by the element $\mathrm{S}\mathrm{T}^{-1}$, i.e.\ working in the limit
$\epsilon \rightarrow 0^+$,
\begin{equation}
\mathrm{S}\mathrm{T}^{-1} \,\circ\, \vev T = \begin{pmatrix} 0 & 1\\ -1 & 1\end{pmatrix}\circ\, \vev T 
~=~ \frac{0\,\vev T + 1}{-1\,\vev T+1} ~\xrightarrow{\epsilon \rightarrow 0^+}~ \I\,\infty 
\quad\mathrm{for}\quad \vev T = 1+\I \epsilon ~\xrightarrow{\epsilon \rightarrow 0^+}~ 1\;.
\end{equation}
To find the local enhancement of the traditional flavor symmetry by the stabilizer of the modulus, 
note that $\vev T = 1$ is left invariant by the $\gamma \in \SL{2,\Z{}}_T$ transformations satisfying
\begin{equation}
\label{eq:StabilizerAtT1}
\gamma \,\circ\, \vev T ~=~ \begin{pmatrix} a & b\\ c & d\end{pmatrix}\circ\, \vev T 
~=~ \frac{a\,\vev T + b}{c\,\vev T + d} ~\stackrel{!}{=}~ \vev T \quad\xLeftrightarrow{~\vev T = 1~}\quad a + b ~\stackrel{!}{=}~ c + d\;.
\end{equation}
Combining this with the constraint $a d - b c = 1$, we obtain 
\begin{equation}
\left(c + d - b\right) d - b c ~=~ 1 \quad\Leftrightarrow\quad \left(c + d\right)\left(d - b\right) ~=~ 1\;.
\end{equation}
Recalling that all variables here are integers, we observe that
\begin{equation}
s~:=~c + d ~=~ d - b \qquad\text{with}\qquad s = \pm 1 \qquad\Rightarrow\qquad \gamma ~=~ \begin{pmatrix}s-b & b \\ -b& s+b\end{pmatrix}\;.
\end{equation}
This yields the stabilizer at $\vev T = 1$. All solutions to 
eq.~\eqref{eq:StabilizerAtT1} then read
\begin{equation}\label{eq:StabilizerAtT1Solutions}
\begin{pmatrix}s-b & b \\ -b& s+b\end{pmatrix} ~=~ \left\{
\begin{array}{ll}
\left(\mathrm{ST^{-2}}\right)^{b}                & \mathrm{for\ } s = +1\;,\\
\mathrm{S}^2\,\left(\mathrm{ST^{-2}}\right)^{-b} & \mathrm{for\ } s = -1\;,
\end{array}
\right.
\end{equation}
for $b \in\Z{}$. Therefore,
the stabilizer can be generated by the two elements
\begin{equation}
\mathrm{S}^2 ~=~ \begin{pmatrix}-1 & 0 \\ 0& -1\end{pmatrix} \quad\mathrm{and}\quad 
\mathrm{ST^{-2}} ~=~ \begin{pmatrix}0 & 1 \\ -1& 2\end{pmatrix}\;.
\end{equation}
The first generator corresponds to the $\Delta(54)$ element $\mathrm{C}$, while the second generator 
\textit{locally} (at $\vev T = 1$) enhances the traditional flavor symmetry $\Delta(54)$ 
to a larger group. Noting that 
$\mathrm{ST}^{-2}=\mathrm{T}\mathrm{S}^{-1}\mathrm{T}\mathrm{S}\mathrm{T}^{-1}=(\mathrm{S}\mathrm{T}^{-1})^{-1}\mathrm{T}(\mathrm{S}\mathrm{T}^{-1})$ 
shows that the second generator is dual to the $\mathrm{T}$ transformation $T\mapsto T+1$ 
that leaves $\vev T = \I\,\infty$ invariant.
At the level of the finite modular group $T'$, the $\SL{2,\Z{}}_T$ transformation 
$\mathrm{ST}^{-2}$ is realized as the $\Z{3}$ transformation $\rho(\mathrm{ST^{-2}}) = \rho(\mathrm{ST})$.
The automorphy factor of $(\mathrm{ST}^{-2})^b$ is trivial at $\vev T = 1$ 
for any modular form of weight $n$, i.e.\ using eq.~\eqref{eq:StabilizerAtT1Solutions} for $s=1$,
we obtain
\begin{equation}
\left(c\,\vev T + d\right)^{n} ~=~ \left(-b+1+b\right)^{n} ~=~ 1\;.
\end{equation}
Hence, for matter fields in $\Delta(54)$ representations $\rep{r} = \rep{3}_2$ 
(for $\Phi_{\nicefrac{-2}{3}}$) or $\rep{r} = \rep{3}_1$ (for $\Phi_{\nicefrac{-5}{3}}$) 
the explicit representation matrix of the generator that locally enhances the traditional 
flavor symmetry is given by
\begin{equation}
\label{eq:Z3atT1}
\rho_{\rep{r},\vev{T}=1}(\mathrm{ST}) ~=~ \rho(\mathrm{ST})\;.
\end{equation}
We note that the $s=-1$ case would give rise to a nontrivial automorphy factor and, 
hence, different matrix generators for $\rep{3}_2$ and 
$\rep{3}_1$, but ultimately lead to exactly the same group.

Hence, altogether, the unified flavor group at $\vev T = 1$ can be generated by
\begin{align}
\label{eq:GenT1Phi23}
 &\Phi_{\nicefrac{-2}{3}}:& 
 \rho_{\rep{3}_2,1}(\mathrm{A})&=\rho(\mathrm A),&
 \rho_{\rep{3}_2,1}(\mathrm{B})&=\rho(\mathrm B),&
 \rho_{\rep{3}_2,1}(\mathrm{C})&=\rho(\mathrm C),& \\\notag
 && 
 \rho_{\rep{3}_2,1}(\mathrm{R})&=\e^{\nicefrac{2\pi\I}{9}}\,\Id_3,&
 \rho_{\rep{3}_2,1}(\mathrm{ST})&=\rho(\mathrm{ST}), &\text{and}  \\
 \label{eq:GenT1Phi53}
 &\Phi_{\nicefrac{-5}{3}}:&  
 \rho_{\rep{3}_1,1}(\mathrm{A})&=\rho(\mathrm A),&
 \rho_{\rep{3}_1,1}(\mathrm{B})&=\rho(\mathrm B),&
 \rho_{\rep{3}_1,1}(\mathrm{C})&=-\rho(\mathrm C),& \\\notag
 &&
 \rho_{\rep{3}_1,1}(\mathrm{R})&=\e^{\nicefrac{-4\pi\I}{9}}\,\Id_3,&
 \rho_{\rep{3}_1,1}(\mathrm{ST})&=\rho(\mathrm{ST}).&
\end{align}
One can show that the groups generated by eqs.~\eqref{eq:GenTwPhi23} and~\eqref{eq:GenT1Phi23}, 
as well as by eqs.~\eqref{eq:GenTwPhi53} and~\eqref{eq:GenT1Phi53} are identical.

To identify the \CP-like stabilizer at $\langle T\rangle=1$, we solve 
\begin{equation}\label{eq:CPStabilizerAtT1}
\gamma_{\CP}\circ \vev T ~=~ \frac{a\,\langle\bar{T}\rangle + b}{c\,\langle\bar{T}\rangle + d} 
~\stackrel{!}{=}~ \vev T \quad\xLeftrightarrow{~\vev T = 1~}\quad a + b ~\stackrel{!}{=}~ c + d\;,
\end{equation}
for an element $\gamma_{\CP} \in \GL{2,\Z{}}_T$ with $\det\gamma_{\CP}=-1$. 
Hence, altogether
\begin{equation}
 \left(c+d\right)\left(d-b\right)\stackrel{!}{=}-1\;.
\end{equation}
Once more, recalling that the matrix representation of $\gamma_{\CP}$ has 
integer entries, we see that
\begin{equation} 
 s~:=~(c+d)~=~-(d-b)\qquad\text{with}\qquad s~=~\pm1 \qquad\Rightarrow\qquad
 \gamma_{\CP} ~=~ \begin{pmatrix}s-b & b \\ 2s-b & -s+b\end{pmatrix}\;.
\end{equation}
This yields the \CP-like stabilizer at $\vev T = 1$. The solutions to eq.~\eqref{eq:CPStabilizerAtT1} can be written as
\begin{equation}\label{eq:CPStabilizerAtT1Solutions}
\begin{pmatrix}s-b & b \\ 2s-b & -s+b\end{pmatrix} ~=~ \left\{
\begin{array}{ll}
 \left(\mathrm{S\,T}^{-2}\right)^{-b+1}\,\mathrm{S\,K}_*& \mathrm{for\ } s~=~+1\;,\\
\mathrm{S}^2\,\left(\mathrm{S\,T}^{-2}\right)^{b+1}\,\mathrm{S\,K}_* & \mathrm{for\ } s~=~-1\;,
\end{array}
\right.
\end{equation}
for $b \in\Z{}$. Since all these transformations differ only by an element of the stabilizer group itself,
it is clear that they give rise to equivalent resulting groups. 
For definiteness, we choose for the generator of the \CP-like stabilizer the element with
$s=1$ and $b=2$, i.e.\ $\rho(\left(\mathrm{ST}^{-2}\right)^{-1}\mathrm{S\,K}_*)=\rho(\mathrm{T}^2\mathrm{K}_*)$, 
such that the resulting $\Z2^\CP$ acts on the modulus and matter fields as
\begin{align}\label{eq:CPtrafoT1}
 &T\xmapsto{\CP}-\bar{T}+2\;,& &\text{and}& &\Phi_n\xmapsto{\CP}\rho(\mathrm{T}^2)\bar{\Phi}_n\;.&
\end{align}
Combining this with the unified flavor group, the full linearly realized unified flavor group at $\langle T\rangle=1$ results as
\begin{equation}
\label{eq:UnifiedFlavorGroupT=1}
\Delta(54) ~\cup~ \Z{9}^R ~\cup~ \mathrm{S}\mathrm{T} ~\cup~ \Z2^\CP 
~=~ H(3,2,1)\rtimes\Z2^\CP ~\cong~ [486,125]\rtimes\Z2~\cong~ [972,469]\;.
\end{equation}
We realize that this result at $\vev T=1$ coincides with the linearly realized unified 
flavor group at $\vev T=\omega$, eq.~\eqref{eq:UnifiedFlavorGroupT=w}.

For completeness, we note that the generator of the \CP-like stabilizer at the dual point 
$\vev T = \I\,\infty$ can be represented by $\mathrm{S}^2\mathrm{K}_*=-\mathrm{K}_*$,
which acts as $T\mapsto-\bar{T}$ and clearly leaves $\vev T = \I\,\infty$ invariant. 
The resulting linearized flavor group at $\vev T = \I\,\infty$ is, as expected, 
the same as for $\vev T = 1$.

\subsection[Heterotic configurations of T6/Z3xZ3 with Omega(2) eclectic symmetry]{\boldmath Heterotic configurations of $\mathbb T^6/(\Z3\x\Z3)$ with $\Omega(2)$ eclectic symmetry \unboldmath}
\label{sec:T6overZ3xZ3models}

\begin{table}[t]
	\centering
	\resizebox{\textwidth}{!}{ 
		\begin{tabular}{clllllllll}
			\toprule
			Model & $\ell$                   & $\bar e$              & $\bar\nu$             & $q$                   & $\bar u$
			& $\bar d$              & $H_u$                 & $H_d$                 & flavons \\
			\midrule
			A & $\Phi_{\nicefrac{-2}3}$ & $\Phi_{\nicefrac{-2}3}$ & $\Phi_{\nicefrac{-2}3}$ & $\Phi_{\nicefrac{-2}3}$ & $\Phi_{\nicefrac{-2}3}$
			& $\Phi_{\nicefrac{-2}3}$ & $\Phi_{0}$  & $\Phi_{0}$  & $\Phi_{\nicefrac{-2}3,-1}$ \\
			B & $\Phi_{\nicefrac{-1}3}$ & $\Phi_{\nicefrac{-2}3}$ & $\Phi_{\nicefrac{-2}3}$ & $\Phi_{\nicefrac{-2}3}$ & $\Phi_{\nicefrac{-2}3}$
			& $\Phi_{\nicefrac{-1}3}$ & $\Phi_{-1}$ & $\Phi_{0}$  & $\Phi_{\nicefrac{-2}3,-1}$ \\
			C & $\Phi_{\nicefrac{-2}3}$ & $\Phi_{\nicefrac{-1}3}$ & $\Phi_{\nicefrac{-1}3}$ & $\Phi_{\nicefrac{-1}3}$ & $\Phi_{\nicefrac{-1}3}$
			& $\Phi_{\nicefrac{-2}3}$ & $\Phi_{-1}$ & $\Phi_{-1}$ & $\Phi_{\nicefrac{-1}3,-1}$\\
			D & $\Phi_{\nicefrac{-1}3}$ & $\Phi_{\nicefrac{-1}3}$ & $\Phi_{\nicefrac{\pm2}3,0}$ & $\Phi_{\nicefrac{-1}3}$ & $\Phi_{\nicefrac{-1}3}$
			& $\Phi_{\nicefrac{-1}3}$ & $\Phi_{0}$ & $\Phi_{-1,0}$ & $\Phi_{\nicefrac{\pm2}3,-1}$\\
			E & $\Phi_{\nicefrac{-2}3,\nicefrac{-1}3}$ & $\Phi_{\nicefrac{-2}3,0}$ & $\Phi_{0,\nicefrac{-2}3,\nicefrac{-1}3,\nicefrac{-5}3}$ & $\Phi_{-1,\nicefrac{-2}3}$ & $\Phi_{\nicefrac{-2}3}$
			& $\Phi_{0,\nicefrac{-2}3}$ & $\Phi_{0}$ & $\Phi_{0}$ & $\Phi_{\nicefrac{-2}3,\nicefrac{-1}3,\nicefrac{-5}3,-1}$\\
			\bottomrule
		\end{tabular}
	}
	\caption{\label{tab:Z3xZ3configurations}
		Flavor symmetry representations of MSSM matter fields in $\mathbbm T^2/\Z3$ orbifold 
		sectors of five different types of six-dimensional $\mathbbm T^6/(\Z3\x\Z3)$ heterotic 
		orbifold models (consistent string theory configurations, see text). 
		For each different type of configuration, we display \textit{all} possibilities for representations 
		that multiplets of quark ($q,\bar u,\bar d$), lepton ($\ell,\bar e,\bar\nu$), and Higgs superfields 
		can take on in the relevant $\mathbbm T^2/\Z3$ orbifold sector. 
		We use the field notation of table~\ref{tab:Representations}, where the subindices denote modular weights. 
		Multiple subindices indicate that matter fields of all those modular weights appear in the respective model.}
\end{table}

The $\mathbbm T^2/\Z3$ orbifold sector appears naturally in six-dimensional $\mathbbm T^6/\Z3$, 
$\mathbbm T^6/\Z6$-II, $\mathbbm T^6/(\Z3\x\Z3)$ and $\mathbbm T^6/(\Z3\x\Z6)$ orbifolds. These
orbifolds have been explored in the search of models arising from heterotic string 
compactifications with the MSSM spectrum plus vectorlike exotics~\cite{Nilles:2014owa,Olguin-Trejo:2018wpw,Parr:2020oar}. 
The details of the matter spectra of these constructions are determined by the choices of the
embedding of the six-dimensional orbifold into the gauge degrees of freedom, subject to 
consistency requirements, such as modular invariance, see 
e.g.~\cite{Bailin:1999nk,Ploger:2007iq,Vaudrevange:2008sm,Ramos-Sanchez:2008nwx}. 

Inspecting the consistent semirealistic string models classified in~\cite{Nilles:2014owa,Olguin-Trejo:2018wpw,Parr:2020oar}
endowed with a $\mathbbm T^2/\Z3$ orbifold sector, we observe that the light MSSM matter 
superfields appear only in a reduced number of field configurations. For example, considering 
the $\mathbbm T^6/(\Z3\x\Z3)$ (1,1) orbifold geometry (see ref.~\cite{Fischer:2012qj} for 
details on this geometry) with one and two vanishing Wilson lines, we identify five types 
of configurations of massless MSSM matter superfields, as summarized in
table~\ref{tab:Z3xZ3configurations} in terms of the field labels used in 
table~\ref{tab:Representations}. Given these configurations, one can arrive at the flavor 
phenomenology of the $\Omega(2)$ eclectic flavor symmetry by using the effective superpotential
and K\"ahler potential given in ref.~\cite{Nilles:2020kgo}, including the spontaneous breakdown
of the eclectic flavor group triggered by the vevs of the indicated flavon representations and
the K\"ahler modulus $T$. This shall be done explicitly for models of configuration type A in 
a companion paper~\cite{Baur:2021pr}.

To conclude this section, let us stress an important empirical TD observation.
As we see in table~\ref{tab:Representations} (and use in table~\ref{tab:Z3xZ3configurations}), 
the modular weights alone suffice to characterize the transformation behavior of fields under 
the flavor symmetries. That is, there is here a one-to-one relation between the modular weights and 
all flavor symmetry charges of the fields. This also holds for other known TD constructions, 
see e.g.~\cite{Kikuchi:2021ogn,Baur:2020jwc,Baur:2021mtl,Almumin:2021fbk,Ishiguro:2021ccl}. If this feature 
turned out to be correct for generic TD models, it would suggest that consistent BU constructions 
should abide to the same rule. Namely, fields of the same modular weight should also transform 
in the very same representation of all modular flavor symmetries.

\section{Breaking of the traditional flavor symmetry \texorpdfstring{$\boldsymbol{\Delta(54)}$}{Delta(54)}}
\label{sec:Delta54Breaking}

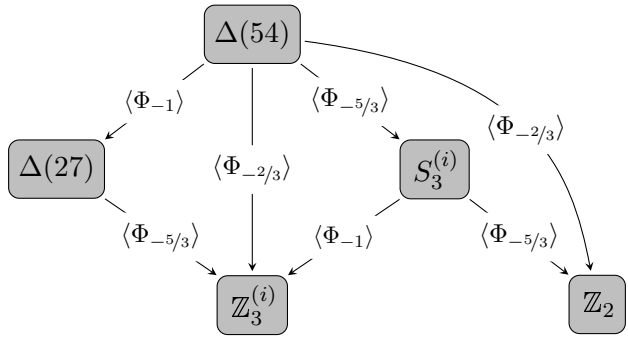
\begin{figure}[!t!]
	\centering
	\begin{tikzpicture}[node distance=1cm and 1.3cm, rounded corners, >=stealth]
	
	\node[minimum height=22pt, draw, rectangle,fill=lightgray] (D54) {$\Delta(54)$ };
	\node[minimum height=22pt, draw, rectangle, below left=of D54,fill=lightgray] (D27) { $\Delta(27)$ };
	\node[minimum height=22pt, draw, rectangle, below right=of D54,fill=lightgray] (S3) { $\;\!S_3^{(i)}$ };
	\node[minimum height=22pt, draw, rectangle, below = 2.8cm of D54,fill=lightgray] (Z3) { $\;\!\Z3^{(i)}$ };
	\node[minimum height=22pt, draw, rectangle, below right=of S3,fill=lightgray] (Z2) { $\;\!\Z2$ };
	
	\draw[->, >=stealth] (D54.south west) -- node [fill=white,rectangle,midway,align=center] {\resizebox*{23pt}{!}{$\vev{\Phi_{-1}}$}} (D27.north east);
	\draw[->, >=stealth] (D54.south east) -- node [fill=white,rectangle,midway,align=center] {\resizebox*{30pt}{!}{$\vev{\Phi_{-\nicefrac53}}$}} (S3.north west);
	\draw[->, >=stealth] (D27.south east) -- node [fill=white,rectangle,midway,align=center] {\resizebox*{30pt}{!}{$\vev{\Phi_{-\nicefrac53}}$}} (Z3.north west);
	\draw[->, >=stealth] (S3.south west) -- node [fill=white,rectangle,midway,align=center] {\resizebox*{23pt}{!}{$\vev{\Phi_{-1}}$}}(Z3.north east);
	\draw[->, >=stealth] (S3.south east) -- node [fill=white,rectangle,midway,align=center] {\resizebox*{30pt}{!}{$\vev{\Phi_{-\nicefrac53}}$}} (Z2.north west);
	\draw[->, >=stealth, shorten >=1.5pt, shorten <=1.5pt] (D54.south) -- node [fill=white,rectangle,midway,align=center] {\resizebox*{30pt}{!}{$\vev{\Phi_{-\nicefrac23}}$}} (Z3.north);
	\draw[->, shorten >=1.5pt, shorten <=1.5pt] (D54) .. controls ($(S3)+(1.3,1.3)$)  and ($(S3)+(1.8,0)$) .. node [fill=white,rectangle,midway,align=center] {\resizebox*{30pt}{!}{$\vev{\Phi_{-\nicefrac23}}$}} (Z2);
	\end{tikzpicture}
	\caption{\label{fig:D54breaking}
		Spontaneous breakdown patterns of the $\Delta(54)$ traditional flavor symmetry 
		of the $\mathbbm T^2/\Z3$ orbifold sector due to different flavon vevs. The index $i=1,2,3,4$ 
		in $S_3^{(i)}$ and $\Z3^{(i)}$ labels various (nonconjugate) subgroups 
		built by different generators, see table~\ref{tab:D54breaking}.}
\end{figure}

In order to understand the breakdown of the $\Omega(2)$ eclectic flavor symmetry of 
the $\mathbbm T^2/\Z3$ orbifold sector, let us first discuss the breaking of the 
traditional flavor symmetry $\Delta(54)\subset\Omega(2)$. Since $\Delta(54)$ is universal 
in moduli space, it can only be broken via nontrivial vevs of flavon fields, but not by the 
moduli. Motivated by the spectra identified in  section \ref{sec:T6overZ3xZ3models}, we consider 
the three matter multiplets $\Phi_{-1}$, $\Phi_{\nicefrac{-2}{3}}$, and $\Phi_{\nicefrac{-5}{3}}$ 
as possible candidates for flavon fields. Depending on their structure, the vevs can either 
preserve different subgroups of $\Delta(54)$ or break the traditional 
flavor symmetry of the $\mathbbm T^2/\Z3$ orbifold sector completely.
The possible breaking patterns of individual vevs are schematically displayed in 
figure~\ref{fig:D54breaking}.

Let us consider a specific $\Delta(54)\to S_3$ example to understand how to arrive at the 
details of the breakdown patterns. The $\Delta(54)$ elements $\mathrm{A}$ and $\mathrm{C}$ 
generate a traditional flavor subgroup $S_3\subset\Delta(54)$ that we denote as $S_3^{(1)}$.
This group corresponds to the geometric permutation symmetry of the $\mathbbm T^2/\Z3$ orbifold 
sector. Any representation of $\Delta(54)$ can be decomposed into irreducible representations 
of this subgroup. If the decomposition of a particular representation includes one or more trivial 
singlets of $S_3^{(1)}$, then the spontaneous breakdown of $\Delta(54)$ to $S_3^{(1)}$ occurs if 
a flavon field transforming in such a particular representation develops a vev along the trivial 
singlet direction(s). Using the characters of $\mathrm{A}$ and $\mathrm{C}$, it is straightforward 
to determine how the $\Delta(54)$ representations of flavon candidates branch in $S_3^{(1)}$ (see 
e.g.\ \cite{Ishimori:2010au} for a pedagogical introduction):
\begin{subequations}
\begin{align}
\vev{\Phi_{-1}}             ~&:~ \rep{1}' \;\!~\rightarrow~ \rep{1}'\;, \\
\vev{\Phi_{\nicefrac{-2}3}} ~&:~ \rep{3}_2 ~\rightarrow~ \rep{1}' \oplus \rep{2}\;, \\
\vev{\Phi_{\nicefrac{-5}3}} ~&:~ \rep{3}_1 ~\rightarrow~ \rep{1} \oplus \rep{2}\;.
\end{align}
\end{subequations}
We see that only the representation $\rep{3}_1$ associated with the matter field $\Phi_{\nicefrac{-5}3}$
branches into a trivial singlet of $S_3$. Therefore, any nonzero vev of $\Phi_{-1}$ or 
$\Phi_{\nicefrac{-2}3}$ breaks not only $\Delta(54)$ but also the $S_3^{(1)}$ subgroup. 
Only the vev of $\Phi_{\nicefrac{-5}3}$ along the direction in field space associated with 
the trivial singlet can result in an unbroken $S_3^{(1)}$ symmetry generated by $\mathrm A$ 
and $\mathrm C$. This direction can be identified in the $\rep3_1$ $\Delta(54)$ representation 
of $\Phi_{\nicefrac{-5}{3}}$ as the simultaneous eigenvector of 
$\rho_{\rep{3}_1}(\mathrm{A})=\rho(\mathrm{A})$ and $\rho_{\rep{3}_1}(\mathrm{C})=-\rho(\mathrm{C})$
with eigenvalue $1$, which is found to be $(1,1,1)^\mathrm{T}$, up to an arbitrary overall factor. 
Note that there are physically equivalent subgroups conjugate to the $S_3^{(1)}$ group generated 
by $\mathrm A$ and $\mathrm C$.
Each choice of conjugate generators na\"ively yields
a different vev structure, but all of them are related by conjugation with an element of $\Delta(54)$.
By contrast, any pattern of $\vev{\Phi_{\nicefrac{-5}3}}$ vevs that cannot be related to $(1,1,1)^\mathrm{T}$
by conjugation will break $S_3^{(1)}$ too.

The previous procedure can be repeated for different choices of generators, which yield different 
subgroups of $\Delta(54)$. All remnant subgroups (up to conjugation), a choice of their corresponding 
generators, and the flavon vev patterns associated with the spontaneous breakdown of $\Delta(54)$ to such 
subgroups are listed in table \ref{tab:D54breaking}.
As in the case of $\Delta(54)\to S_3^{(1)}$, it mostly suffices to consider the vev of a single 
flavon to arrive at the different traditional flavor subgroups. To arrive at the subgroup 
$\Z{3}^{(1)}$, for example, it suffices that $\vev{\Phi_{\nicefrac{-2}3}}$ acquires a nontrivial vev. 
However, if $\Phi_{\nicefrac{-2}3}$ is not present in the spectrum, one may arrive at the same $\Z{3}^{(1)}$ 
subgroup by letting $\Phi_{\nicefrac{-5}3}$ and $\Phi_{-1}$ develop vevs simultaneously.

\begin{table}[t]
	\centering
	\resizebox{\textwidth}{!}{ 
		\begin{tabular}{cccccccc}
			\toprule
			\multirow{2}{*}{$\begin{array}{c} \Delta(54) \\ \text{subgroup} \end{array}$} & \multicolumn{3}{c}{branchings} & \multirow{2}{*}{$\begin{array}{c} \text{subgroup} \\ \text{generator(s)} \end{array}$} & \multicolumn{2}{c}{corresponding vevs} \\ 
			& $\Phi_{-1}$ & $\Phi_{-\nicefrac23}$ & $\Phi_{-\nicefrac53}$ && $\langle\Phi_{-\nicefrac23}\rangle$ & $\langle\Phi_{-\nicefrac53}\rangle$ \\
			\midrule
			$\Delta(27)$ & $\rep{1}\phantom{'}$ & $\rep3$ & $\rep3$ & $\mathrm{A}, \mathrm{B}$  & $(0,0,0)^\mathrm{T}\oplus\vev{\Phi_{-1}}$ & $(0,0,0)^\mathrm{T}\oplus\vev{\Phi_{-1}}$ \\
			\midrule 
			$\mathrm{S}_3^{(1)}$ & $\rep{1}'$ & $\rep1'\oplus\rep2$ & $\rep1\oplus\rep2$ & $\mathrm{A}, \mathrm{C}$  & -- & $(1,1,1)^\mathrm{T}$ \\
			$\Z3^{(1)}$ & $\rep{1}\phantom{'}$ & \multicolumn{2}{c}{$\rep{1}\oplus\rep{1}_\omega\oplus\rep{1}_{\omega^2}$} & $\mathrm{A}$ & $(1,1,1)^\mathrm{T}$ & $(1,1,1)^\mathrm{T}\oplus\langle\Phi_{-1}\rangle$ \\
			\midrule
			$\mathrm{S}_3^{(2)}$ & $\rep{1}'$ & $\rep1'\oplus\rep2$ & $\rep1\oplus\rep2$ & $\mathrm{B}, \mathrm{C}$  & -- & $(1,0,0)^\mathrm{T}$ \\
			$\Z3^{(2)}$ & $\rep{1}\phantom{'}$ & \multicolumn{2}{c}{$\rep{1}\oplus\rep{1}_\omega\oplus\rep{1}_{\omega^2}$} & $\mathrm{B}$ & $(1,0,0)^\mathrm{T}$ & $(1,0,0)^\mathrm{T}\oplus\langle\Phi_{-1}\rangle$ \\
			\midrule
			$\mathrm{S}_3^{(3)}$ & $\rep{1}'$ & $\rep1'\oplus\rep2$ & $\rep1\oplus\rep2$ & $\mathrm{ABA}, \mathrm{C}$  & -- & $(\omega,1,1)^\mathrm{T}$ \\
			$\Z3^{(3)}$ & $\rep{1}\phantom{'}$ & \multicolumn{2}{c}{$\rep{1}\oplus\rep{1}_\omega\oplus\rep{1}_{\omega^2}$} & $\mathrm{ABA}$ & $(\omega,1,1)^\mathrm{T}$ & $(\omega,1,1)^\mathrm{T}\oplus\langle\Phi_{-1}\rangle$ \\
			\midrule
			$\mathrm{S}_3^{(4)}$ & $\rep{1}'$ & $\rep1'\oplus\rep2$ & $\rep1\oplus\rep2$ & $\mathrm{AB^2A}, \mathrm{C}$  & -- & $(\omega^2,1,1)^\mathrm{T}$ \\
			$\Z3^{(4)}$ & $\rep{1}\phantom{'}$ & \multicolumn{2}{c}{$\rep{1}\oplus\rep{1}_\omega\oplus\rep{1}_{\omega^2}$} & $\mathrm{AB^2A}$ & $(\omega^2,1,1)^\mathrm{T}$ & $(\omega^2,1,1)^\mathrm{T}\oplus\vev{\Phi_{-1}}$ \\
			\midrule
			$\Z2$ & $\rep{1}_{-1}$ & $\rep{1}\oplus\rep{1}_{-1}\oplus\rep{1}_{-1}$ & $\rep{1}\oplus\rep{\rep{1}}\oplus\rep{1}_{-1}$ & $\mathrm{C}$ &  $(0,1,-1)^\mathrm{T}$ & 
			$\begin{array}{c} (1,0,0)^\mathrm{T} \\[-2pt] +\alpha(0,1,1)^\mathrm{T} \end{array}$\\
			\bottomrule
		\end{tabular}
	}
	\caption{\label{tab:D54breaking}
		Overview of subgroups of the $\Delta(54)$ traditional flavor group and the corresponding vevs for
		the breakdown. The second column shows the branching of the relevant representations into the subgroups. 
		The $\rep{1}$ here denotes the trivial singlet, and nontrivial singlets are labeled by 
		their eigenvalue under the Abelian generator. We provide examples for generators of each subgroup (up to conjugation) 
		that are left unbroken by the different choices of vevs specified in the last two columns. The notation 
		``$\oplus\,\langle\Phi_{-1}\rangle$'' means that a vev of a nontrivial $\Delta(54)$ singlet field has to be switched 
		on in addition to the vev of a triplet in order to achieve the breaking to the respective subgroup.
		The provided generators and vevs are not unique since their conjugation can yield equivalent $\Delta(54)$ subgroups.
		All the subgroups are stated up to conjugation, i.e.\ the groups are distinct and not related by conjugation. 
		We omit here an arbitrary global (normalization) factor for each of the 
		vevs, $\alpha$ is an arbitrary complex number, while ``--'' denotes the absence of a suitable vev.}
\end{table}

Altogether we see that vevs $\vev{\Phi_{-1}}$, $\vev{\Phi_{\nicefrac{-2}3}}$, and $\vev{\Phi_{\nicefrac{-5}3}}$ 
are quite efficient in breaking the traditional flavor symmetry. However, they can leave remnant symmetries 
unbroken and this might happen by default, as potentials are known to be often minimized at symmetry enhanced 
points (i.e.\ by vevs with remnant subgroups). In order to break the flavor symmetry completely, vevs can 
be misaligned from their symmetry enhanced directions (thereby still allowing for \textit{approximate} or 
softly broken symmetries that can give rise to ``flavor hierarchies''), or multiple vevs could be present simultaneously.
From the stated patterns of single vevs it is straightforward to work out remnant groups also if multiple 
vevs are present. The respective remnant groups would be given as the nontrivial intersection of the 
preserved symmetries of each individual vev. We will further explore the phenomenological consequences 
of these scenarios in our forthcoming paper~\cite{Baur:2021pr}.

\section{Breaking of the complete eclectic flavor symmetry \texorpdfstring{$\boldsymbol{\Omega(2)}$}{Omega(2)}}
\label{sec:Omega2Breaking}

\begin{figure}[t!]
	\centering
	\begin{tikzpicture}[node distance=1cm and 1.3cm, rounded corners, >=stealth]
	
	\node[minimum height=22pt, draw, rectangle,fill=lightgray] (D54) {$\;\!\Xi(2,2)$ };
	\node[minimum height=22pt, draw, rectangle, below left=of D54,fill=lightgray] (D27) { $\Delta(27)$ };
	\node[minimum height=22pt, draw, rectangle, below right=of D54,fill=lightgray] (S3) { $\;\!S_3^{(i)}$ };
	\node[minimum height=22pt, draw, rectangle, below = 2.8cm of D54,fill=lightgray] (Z3) { $\;\!\Z3^{(i)}$ };
	\node[minimum height=22pt, draw, rectangle, below right= of S3,fill=lightgray] (Z2) { $\;\!\Z2$ };
	\node[minimum height=22pt, draw, rectangle, right= 3.4cm of S3,fill=lightgray] (Z4) { $\;\!\Z4$ };
	
	\draw[->, >=stealth] (D54.south west) -- node [fill=white,rectangle,midway,align=center] {\resizebox*{23pt}{!}{$\vev{\Phi_{-1}}$}} (D27.north east);
	\draw[->, >=stealth] (D54.south east) -- node [fill=white,rectangle,midway,align=center] {\resizebox*{30pt}{!}{$\vev{\Phi_{-\nicefrac53}}$}} (S3.north west); 
	\draw[->, >=stealth] (D27.south east) -- node [fill=white,rectangle,midway,align=center] {\resizebox*{30pt}{!}{$\vev{\Phi_{-\nicefrac53}}$}} (Z3.north west);
	\draw[->, >=stealth] (S3.south west) -- node [fill=white,rectangle,midway,align=center] {\resizebox*{23pt}{!}{$\vev{\Phi_{-1}}$}} (Z3.north east);
	\draw[->, >=stealth] (S3.south east) -- node [fill=white,rectangle,midway,align=center] {\resizebox*{30pt}{!}{$\vev{\Phi_{-\nicefrac53}}$}} (Z2.north west);
	\draw[->, >=stealth, shorten >=1.5pt, shorten <=1.5pt] (D54.south) -- node [fill=white,rectangle,midway,align=center] {\resizebox*{30pt}{!}{$\vev{\Phi_{-\nicefrac23}}$}} (Z3.north);
	\draw[->, shorten >=1.5pt, shorten <=1.5pt] (D54) .. controls ($(S3)+(1.4,0.65)$)  and ($(S3)+(2.,-0.45)$) .. node [fill=white,rectangle,pos=0.56,align=center] {~\resizebox*{30pt}{!}{$\vev{\Phi_{-\nicefrac23}}$}} (Z2);
	\draw[->, shorten >=1.5pt, shorten <=1.5pt] (D54) -- node [fill=white,rectangle,midway,align=center] {\resizebox*{30pt}{!}{$\vev{\Phi_{-\nicefrac53}}$}} (Z4.north west);
	\draw[->, shorten >=1.5pt, shorten <=1.5pt] (Z4.south west) -- node [fill=white,rectangle,midway,align=center] {\resizebox*{30pt}{!}{$\vev{\Phi_{-\nicefrac53}}$}} (Z2.north east);
	\end{tikzpicture}
	\caption{\label{fig:Xi22breaking}
		Spontaneous breakdown patterns of the linearly realized unified 
		flavor symmetry $\Xi(2,2)$ at $\vev{T}=\I$ by $\mathbbm T^2/\Z3$ flavon vevs.
		The index $i=1,2$ in $S_3^{(i)}$ and $\Z3^{(i)}$
		labels various (nonconjugate) 
		subgroups built by different generators, see table~\ref{tab:Xi22}.}
\end{figure}
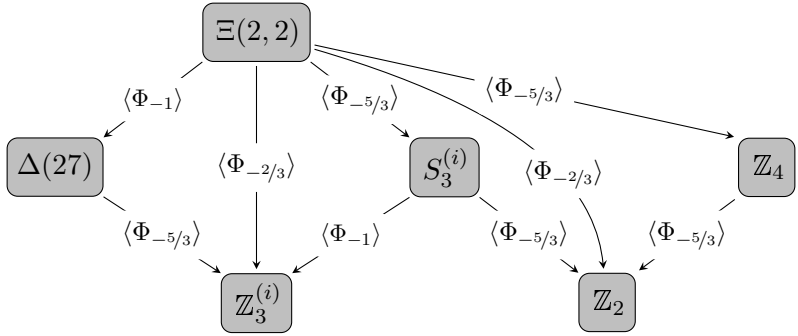

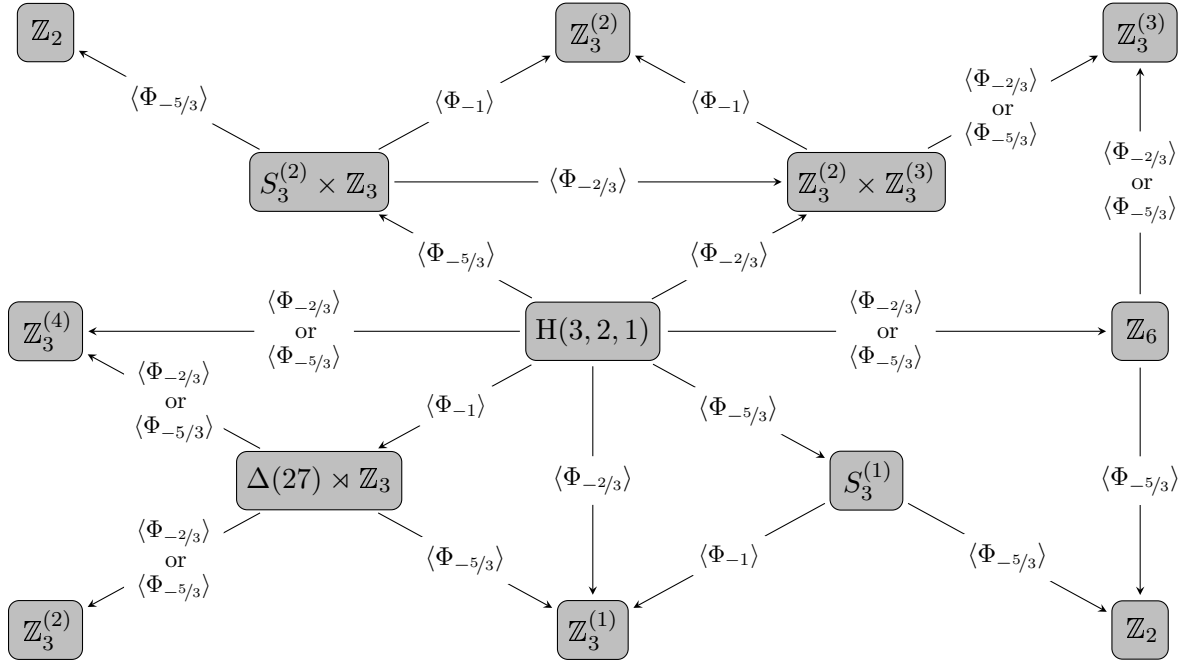
\begin{figure}[t]
	\centering
	\begin{tikzpicture}[node distance=0.7cm and 1cm, rounded corners, >=stealth,
	scale=1.8, shorten >=2, shorten <=2]
	
	\coordinate (H321) at (0,0);
	\coordinate (D27) at (-2,-1.1);
	\coordinate (S31) at (2,-1.1);
	\coordinate (Z62) at (4,0);
	\coordinate (S3xZ3) at (-2,1.1);
	\coordinate (Z3xZ3) at (2,1.1);
	\coordinate (Z31) at (0,-2.2);
	\coordinate (Z2) at (4,-2.2);
	\coordinate (Z32) at (0,2.2);
	\coordinate (Z322) at (-4,-2.2);
	\coordinate (Z33) at (4,2.2);
	\coordinate (Z34) at (-4,0);
	\coordinate (Z35) at (-4,2.2);
	
	\node[minimum height=22pt, draw, rectangle, fill=lightgray] (H321) at (H321) {$\mathrm{H}(3,2,1)$ };
	\node[minimum height=22pt, draw, rectangle, fill=lightgray] (D27) at (D27) {$\Delta(27)\rtimes \Z3$ };
	\node[minimum height=22pt, draw, rectangle, fill=lightgray] (S31) at (S31) {$\;\!S_3^{(1)}$ };
	\node[minimum height=22pt, draw, rectangle, fill=lightgray] (Z62) at (Z62) {$\;\!\Z{6}$ };
	\node[minimum height=22pt, draw, rectangle, fill=lightgray] (S3xZ3) at (S3xZ3) {$S_3^{(2)} \times \Z{3}$ };
	\node[minimum height=22pt, draw, rectangle, fill=lightgray] (Z3xZ3) at (Z3xZ3) {$\Z{3}^{(2)}\times\Z{3}^{(3)}$ };
	\node[minimum height=22pt, draw, rectangle, fill=lightgray] (Z31) at (Z31) {$\Z{3}^{(1)}$ };
	\node[minimum height=22pt, draw, rectangle, fill=lightgray] (Z2) at (Z2) {$\;\!\Z{2}$ };
	\node[minimum height=22pt, draw, rectangle, fill=lightgray] (Z32) at (Z32) {$\;\!\Z{3}^{(2)}$ };
	\node[minimum height=22pt, draw, rectangle, fill=lightgray] (Z322) at (Z322) {$\;\!\Z{3}^{(2)}$ };
	\node[minimum height=22pt, draw, rectangle, fill=lightgray] (Z33) at (Z33) {$\;\!\Z{3}^{(3)}$ };
	\node[minimum height=22pt, draw, rectangle, fill=lightgray] (Z34) at (Z34) {$\;\!\Z{3}^{(4)}$ };
	\node[minimum height=22pt, draw, rectangle, fill=lightgray] (Z35) at (Z35) {$\;\!\Z{2}$ };

	\draw[->, shorten <=3] (H321) -- node [fill=white,rectangle,midway,align=center] {\resizebox*{23pt}{!}{$\vev{\Phi_{-1}}$}} (D27);
	\draw[->, shorten <=2.5] (H321) -- node [fill=white,rectangle,midway,align=center] {\resizebox*{30pt}{!}{$\vev{\Phi_{-\nicefrac53}}$}} (S31);
	\draw[->] (H321) -- node [fill=white,rectangle,midway,align=center] {\resizebox*{30pt}{!}{$\vev{\Phi_{-\nicefrac23}}$}} (Z31);
	
	\draw[->, shorten <=2.8] (H321) -- node [fill=white,rectangle,midway,align=center] {\resizebox*{30pt}{!}{\parbox{36pt}{\centering$\vev{\Phi_{-\nicefrac23}}$\\[-3pt]\,or\\[-5pt]$\vev{\Phi_{-\nicefrac53}}$}}} (Z62);

	\draw[->, shorten <=2.8] (H321) -- node [fill=white,rectangle,midway,align=center] {\resizebox*{30pt}{!}{$\vev{\Phi_{-\nicefrac53}}$}} (S3xZ3);

	\draw[->] (H321) -- node [fill=white,rectangle,midway,align=center] {\resizebox*{30pt}{!}{$\vev{\Phi_{-\nicefrac23}}$}} (Z3xZ3);
	\draw[->] (S3xZ3) -- node [fill=white,rectangle,midway,align=center] {\resizebox*{30pt}{!}{$\vev{\Phi_{-\nicefrac23}}$}} (Z3xZ3);
	\draw[->] (D27) -- node [fill=white,rectangle,midway,align=center] {\resizebox*{30pt}{!}{\parbox{36pt}{\centering$\vev{\Phi_{-\nicefrac23}}$\\[-3pt]\,or\\[-5pt]$\vev{\Phi_{-5/3}}$}}} (Z34);
	\draw[->] (D27) -- node [fill=white,rectangle,midway,align=center] {\resizebox*{30pt}{!}{$\vev{\Phi_{-\nicefrac53}}$}} (Z31);
	\draw[->] (S31) -- node [fill=white,rectangle,midway,align=center] {\resizebox*{23pt}{!}{$\vev{\Phi_{-1}}$}} (Z31);
	\draw[->] (S31) -- node [fill=white,rectangle,midway,align=center] {\resizebox*{30pt}{!}{$\vev{\Phi_{-\nicefrac53}}$}} (Z2);
	\draw[->, shorten <=2.5] (Z62) -- node [fill=white,rectangle,midway,align=center] {\resizebox*{30pt}{!}{$\vev{\Phi_{-\nicefrac53}}$}} (Z2);
	
	\draw[->, shorten <=2.8] (Z62) -- node [fill=white,rectangle,midway,align=center] {\resizebox*{30pt}{!}{\parbox{36pt}{\centering$\vev{\Phi_{-\nicefrac23}}$\\[-3pt]\,or\\[-5pt]$\vev{\Phi_{-\nicefrac53}}$}}} (Z33);
	
	\draw[->, shorten <=2.8] (Z3xZ3) -- node [fill=white,rectangle,pos=0.44,align=center] {\resizebox*{30pt}{!}{\parbox{36pt}{\centering$\vev{\Phi_{-\nicefrac23}}$\\[-3pt]\,or\\[-5pt]$\vev{\Phi_{-\nicefrac53}}$}}} (Z33);

	\draw[->] (H321) -- node [fill=white,rectangle,midway,align=center] {\resizebox*{30pt}{!}{\parbox{36pt}{\centering$\vev{\Phi_{-\nicefrac23}}$\\[-3pt]\,or\\[-5pt]$\vev{\Phi_{-\nicefrac53}}$}}} (Z34);
	
	\draw[->] (S3xZ3) -- node [fill=white,rectangle,midway,align=center] {\resizebox*{30pt}{!}{$\vev{\Phi_{-\nicefrac53}}$}} (Z35);
	
	\draw[->] (S3xZ3) -- node [fill=white,rectangle,midway,align=center] {\resizebox*{23pt}{!}{$\vev{\Phi_{-1}}$}} (Z32);
	
	\draw[->] (Z3xZ3) -- node [fill=white,rectangle,midway,align=center] {\resizebox*{23pt}{!}{$\vev{\Phi_{-1}}$}} (Z32);
	
	\draw[->] (D27) -- node [fill=white,rectangle,midway,align=center] {\resizebox*{30pt}{!}{\parbox{36pt}{\centering$\vev{\Phi_{-\nicefrac23}}$\\[-3pt]\,or\\[-5pt]$\vev{\Phi_{-\nicefrac53}}$}}} (Z322);
	
	\end{tikzpicture}
	\caption{\label{fig:H321breaking}
		Spontaneous breakdown patterns of the linearly realized unified flavor symmetry 
		$\mathrm{H}(3,2,1)$ at $\vev{T}=\omega,1,\I\infty$.
		Here $\Delta(27)\rtimes \Z3$ stands for the group $[81,9]$ when $\vev{T}=\omega$ and 
                for $[81,7]$ in case of $\vev{T}=1,\I\infty$. The two boxes with $\Z2$ and $\Z3^{(2)}$
                represent the same $\Z2$ and $\Z3^{(2)}$ groups, respectively. The various $\Z3^{(i)}$ 
				and $S_3^{(i)}$ are not related by conjugation and have different generators,
                see table~\ref{tab:H321}.
            }
\end{figure}

After the $\Delta(54)$ example, let us consider the spontaneous breakdown patterns 
of the complete eclectic flavor symmetry $\Omega(2)\rtimes\Z2^\CP$. As discussed above, 
only parts of this group can be linearly realized on the spectrum of twisted matter fields,
while other parts are necessarily broken by the vev $\vev T$ of the K\"ahler modulus.
Compatibility with observations requires that, on top of the breaking induced by $\vev T$,
the linearly realized unified flavor symmetry must be further broken by vevs of flavon fields.

At a generic point in moduli space, the traditional flavor symmetry is $\Delta(54)$ and 
the results of the previous section apply. In this section, we investigate 
analogous breaking patterns at the symmetry enhanced points in moduli space. In detail, we study 
the points $\vev T=\I$ and $\vev T=\omega,1,\I\infty$, where the traditional flavor 
symmetry $\Delta(54)$ is enhanced, respectively, to $\Xi(2,2)\cong[324,111]$ 
and $H(3,2,1)\cong[486,125]$ as discussed in sections~\ref{sec:FlavorEnhancementT=i} to \ref{sec:FlavorEnhancementT=1}.
As derived above, these flavor symmetries are additionally enhanced by \CP-like transformations.
However, for practical reasons, we first focus on non-\CP-like transformations
and comment on the possible enhancements by \CP-like transformations in the end. 

We focus on the matter fields $\Phi_{\nicefrac{-2}{3}}$ and $\Phi_{\nicefrac{-5}{3}}$ 
from the $\theta$ sector of a $\mathbbm T^2/\Z3$ orbifold, as well as on the bulk field $\Phi_{-1}$, because these fields arise as 
potential flavons in the models under consideration, see section~\ref{sec:T6overZ3xZ3models}. 
The complete charge assignment of these fields together with the generators under the relevant modular 
and flavor transformations is summarized in table~\ref{tab:Representations}. 
As already stressed at the end of section~\ref{sec:T6overZ3xZ3models}, the complete transformation behavior of a field can 
be inferred already from the respective modular weights which are fixed by the string theory construction.

Our results for the breakdown patterns induced by flavon vevs $\vev{\Phi_{\nicefrac{-2}{3}}}$, 
$\vev{\Phi_{\nicefrac{-5}{3}}}$ and/or $\vev{\Phi_{-1}}$ are depicted in figure~\ref{fig:Xi22breaking} 
for the linearly realized unified flavor symmetry $\Xi(2,2)$ at $\vev T=\I$, and in figure~\ref{fig:H321breaking} 
for the unified flavor symmetry $H(3,2,1)$ at $\vev T=\omega,1,\I\infty$. Details of our results are summarized 
in table~\ref{tab:Xi22} (for $\vev T=\I$) and table~\ref{tab:H321} (for $\vev T=\omega,1,\I\infty$).
Just as in the previous section, we determine the branching of the respective irreducible representations into 
subgroups of the flavor symmetries. If at least one trivial singlet $\rep1$ is present in the branching, the 
associated flavon vev can break the original unified flavor group to the listed subgroup. 
We do not list subgroups that cannot be realized by any of the given vevs because they
would actually preserve a larger subgroup.
We give explicit generators for all subgroups and the associated explicit vevs (up to conjugation).
There exist multiple equivalent subgroups which can be obtained from the stated examples 
by conjugation.

As noted already above, the flavor symmetry enhancement at $\vev T=\omega$, $\vev T=1$ and at $\vev T=\I\infty$ 
leads to isomorphic unified flavor groups. We have confirmed explicitly that also the respective matrix 
groups for the triplets generate identical groups (and irreducible representations). Hence, the analysis 
of the triplet vevs is exactly the same at all three points. For definiteness, we chose in table~\ref{tab:H321} 
the generator convention at $\vev T=\omega$ provided explicitly in  eqs.~\eqref{eq:GenTwPhi23} and~\eqref{eq:GenTwPhi53}.

Here, we comment on the breakdown induced by the vev of the nontrivial singlet field $\vev{\Phi_{-1}}$. 
The transformation behavior of $\Phi_{-1}$ can be read off from table~\ref{tab:Representations}. 
Generators of the remnant modular group $\mathrm{S}$ and $\mathrm{T}$ furthermore have to be amended 
by the correct automorphy factors, stated in \eqref{eq:automorphyI} and~\eqref{eq:automorphyOmega} 
taken for modular weight $n=-1$.
Altogether, we find that the single vev $\langle\Phi_{-1}\rangle$ induces the breakings
\begin{align}
&\langle\Phi_{-1}\rangle:&
&\Xi(2,2)\rightarrow\Delta(27)\;, &
&H(3,2,1)_{\langle T\rangle=\omega}\rightarrow[81,9]\;,  &
&H(3,2,1)_{\langle T\rangle=1,\I\infty}\rightarrow[81,7]\;.  &  
\end{align}
We see that, unlike the triplet winding states, the breaking induced by the vev of the nontrivial 
singlet bulk field $\vev{\Phi_{-1}}$ differs between points $\vev T=\omega$ and $\vev T=1,\I\infty$. 
However, the structure of the groups $[81,9]$ and $[81,7]$ coincides in both cases with $\Delta(27)\rtimes\Z3$,
which is the notation that we adopt in figure~\ref{fig:H321breaking}.

\begin{table}[!t!]
	\centering
\resizebox{\textwidth}{!}{ 
	\begin{tabular}{cccccccc}
		\toprule
		\multirow{2}{*}{$\begin{array}{c} \Xi(2,2) \\ \text{subgroup} \end{array}$} & \multicolumn{2}{c}{branchings} & \multirow{2}{*}{$\begin{array}{c} \text{subgroup} \\ \text{generator(s)} \end{array}$} & \multicolumn{2}{c}{corresponding vevs} \\ 
		                    & $\Phi_{\nicefrac{-2}3}$ & $\Phi_{\nicefrac{-5}3}$ && $\vev{\Phi_{\nicefrac{-2}3}}$ & $\vev{\Phi_{\nicefrac{-5}3}}$ \\
		\midrule
		$\mathrm{S}_3^{(1)}$ & $\rep1'\oplus\rep2$ & $\rep{1} \oplus \rep{2}$ & $\mathrm{A}, \mathrm{C}$  & -- & $(1,1,1)^\mathrm{T}$ \\
		$\Z3^{(1)}$ & $\rep{1}\oplus\rep{1}_\omega\oplus\rep{1}_{\omega^2}$ & $\rep{1}\oplus\rep{1}_\omega\oplus\rep{1}_{\omega^2}$ & $\mathrm{A}$ & $(1,1,1)^\mathrm{T}$ & $(1,1,1)^\mathrm{T}\oplus \vev{\Phi_{-1}}$ \\
		\midrule
		$\mathrm{S}_3^{(2)}$ & $\rep1'\oplus\rep2$ & $\rep{1} \oplus \rep{2}$ & $\mathrm{ABA},\mathrm{C}$ & -- & $(\omega,1,1)^\mathrm{T}$ \\
		$\Z3^{(2)}$ & $\rep{1}\oplus\rep{1}_\omega\oplus\rep{1}_{\omega^2}$ & $\rep{1}\oplus\rep{1}_\omega\oplus\rep{1}_{\omega^2}$ & $\mathrm{ABA}$ & $(\omega,1,1)^\mathrm{T}$ & $(\omega,1,1)^\mathrm{T}\oplus \vev{\Phi_{-1}}$ \\
		\midrule
		$\Z4$ & $\rep{1}_{-1} \oplus \rep{1}_{\I} \oplus \rep{1}_{-\I}$ & $\rep{1}\oplus\rep{\rep1}_{-1}\oplus\rep{\rep1}_\I$ & $\mathrm{S^3}$
		 & -- & $(1+\sqrt{3},1,1)^\mathrm{T}$ \\  
		$\Z2$ & $\rep{1}\oplus\rep{1}_{-1}\oplus\rep{1}_{-1}$ & $\rep{1}\oplus\rep{\rep{1}}\oplus\rep{1}_{-1}$ & $\mathrm{C}$ &  $(0,1,-1)^\mathrm{T}$ & 
		$\begin{array}{c} (1,0,0)^\mathrm{T} \\[-2pt] +\alpha(0,1,1)^\mathrm{T} \end{array}$\\
		\bottomrule
	\end{tabular}
}
	\caption{\label{tab:Xi22}
	Maximal subgroups (up to conjugation) of $\Xi(2,2)$ that can be achieved from its breakdown by vevs of flavon fields
	at $\vev T=\I$. We provide the branchings of the flavon representations under the resulting subgroups, followed 
	by samples of the generators of such subgroups and the flavon vevs that yield the corresponding breakdown. 
	Note that the \Z2 generator satisfies $\mathrm{C}=(\mathrm{S}^3)^2$.
	We follow the same notation as in table~\ref{tab:D54breaking}.
        }
\end{table}

To conclude, let us discuss the stabilizers of the \CP-like type, i.e.\ the possibility of 
enhanced \CP-like transformations.  
Whether or not \CP-like transformations are broken or preserved is often a model dependent 
statement. As we will see, this is because rephasing transformations of fields are important 
in this context, and whether or not those rephasings are physical or unphysical depends on 
the specifics of the model. This is why in our previous discussion we have focused on the 
non-\CP-like flavor symmetries, where model independent statements are possible. However, 
independently of the model one can answer the question of whether or not the vev of a specific field
would automatically break the associated \CP-like transformation (i.e.\ whether or not the 
respective vev has a \CP-like stabilizer). 
The transformations we investigate here are described by eq.~\eqref{eq:CPTrafoFields}, i.e.\
they transform $\Phi\mapsto U\bar{\Phi}$ in short. In addition, the transformation might include 
an element of the initial (non-\CP-like) eclectic flavor symmetry that is preserved at the specific 
location $\vev T$, i.e.\ a transformation of the type 
$\Phi\mapsto \rho(\mathrm{g})U\bar{\Phi}$ with $g\in G_\mathrm{traditional}\,\cup\,G_\mathrm{modular}\,\cup\,G_\mathrm{R}$.
Note that there is a crucial difference here with respect to generic bottom-up constructions: From the BU perspective, 
$U$ has to be a representation matrix of a suitable outer automorphism transformation of the flavor 
symmetry that maps the representations to their complex conjugates. Such matrices are only defined 
up to a global phase by construction. Therefore, in solving the equation 
$U\bar{\Phi}\stackrel{!}{=}\Phi$, a phase can always be absorbed in $U$. 
This is to be contrasted with the TD construction. Here, first of all, not all outer automorphisms 
of the flavor symmetry are possible, but only those which are part of the eclectic group and preserved 
by the specific vev $\vev T$ of the modulus. Second, as the matrix $U$ itself is a representation 
matrix within the eclectic group, arbitrary global phases for $U$ are not admissible.

\begin{table}[!t!]
	\centering
	\resizebox{\textwidth}{!}{ 
		\begin{tabular}{cccccccc}
			\toprule
			\multirow{2}{*}{$\begin{array}{c} H(3,2,1) \\ \text{subgroup} \end{array}$} & \multicolumn{2}{c}{branchings} & \multirow{2}{*}{$\begin{array}{c} \text{subgroup} \\ \text{generator(s)} \end{array}$} & \multicolumn{2}{c}{corresponding vevs} \\ 
			& $\Phi_{\nicefrac{-2}3}$ & $\Phi_{\nicefrac{-5}3}$ && $\vev{\Phi_{\nicefrac{-2}3}}$ & $\vev{\Phi_{\nicefrac{-5}3}}$ \\
			\midrule
			$\mathrm{S}^{(2)}_3\times\Z3$ & $\rep1'_{1}\oplus\rep2_{\omega}$ & $\rep{1}_1\oplus\rep{2}_{\omega}$ & $\mathrm{AC},\mathrm{B^2A^2},\mathrm{R(ST)^4}$ &  -- & $(1,\omega^2,1)^\mathrm{T}$ \\
			$\Z3^{(2)}\times\Z3^{(3)}$ & $\rep{1}\oplus\rep{1}_{\omega^2\!,1}\oplus\rep{1}_{\omega\!,\omega^2}$ & $\rep{1}\oplus\rep{1}_{\omega^2\!,1}\oplus\rep{1}_{\omega,\omega^2}$ & $\mathrm{B^2A^2},\mathrm{R(ST)^4}$ & $(1,\omega^2,1)^\mathrm{T}$ & 
			$\begin{array}{c} (1,\omega^2,1)^\mathrm{T} \\[-2pt] \text{(preserves $\mathrm{S}^{(2)}_3\times\Z3$)} \end{array}$
			\\
			$\Z3^{(2)}$ & $\rep{1}\oplus\rep{1}_{\omega^2}\oplus\rep{1}_{\omega}$ & $\rep{1}\oplus\rep{1}_{\omega^2}\oplus\rep{1}_{\omega}$ & $\mathrm{B^2A^2}$ & $(1,\omega^2,1)^\mathrm{T}\oplus \vev{\Phi_{-1}}$ & $(1,\omega^2,1)^\mathrm{T}\oplus \vev{\Phi_{-1}}$ \\
			$\Z3^{(3)}$ & $\rep{1}\oplus\rep{1}\oplus\rep{1}_{\omega^2}$ & $\rep{1}\oplus\rep{1}\oplus\rep{1}_{\omega^2}$ & $\mathrm{R(ST)^4}$ & \hspace{-0.35cm}$\begin{array}{c} (-\omega^2,1,0)^\mathrm{T} \\[-2pt] +\alpha(-\omega^2,0,1)^\mathrm{T} \end{array}$\hspace{-0.35cm} & \hspace{-0.35cm}$\begin{array}{c}(-\omega^2,1,0)^\mathrm{T} \\[-2pt] +\alpha(-\omega^2,0,1)^\mathrm{T} \end{array}$\hspace{-0.35cm}\\
			\midrule
			$\mathrm{S}_3^{(1)}$ & $\rep1'\oplus\rep2$ & $\rep{1}\oplus\rep{2}$ & $\mathrm{C},\mathrm{A}$ &  -- & $(1,1,1)^\mathrm{T}$ \\
			$\Z{3}^{(1)}$ & $\rep{1}\oplus\rep{1}_\omega\oplus\rep{1}_{\omega^2}$ & $\rep{1}\oplus\rep{1}_\omega\oplus\rep{1}_{\omega^2}$ & $\mathrm{A}$ &  $(1,1,1)^\mathrm{T}$ & $(1,1,1)^\mathrm{T}\oplus \vev{\Phi_{-1}}$ \\
			\midrule
			$\Z{6}$ & $\rep{1}\oplus\rep{1}_{-1}\oplus\rep{1}_{-\omega}$ & $\rep{1}\oplus\rep{1}_{-1}\oplus\rep{1}_{\omega}$ & $\mathrm{CR^2(ST)^8}$ &  $(0,1,-1)^\mathrm{T}$ & $(-2\omega^2,1,1)^\mathrm{T}$ \\
			$\Z{3}^{(3)}$ & $\rep{1}\oplus\rep{1}\oplus\rep{1}_{\omega^2}$ & $\rep{1}\oplus\rep{1}\oplus\rep{1}_{\omega^2}$ & $\mathrm{R(ST)^4}$ &  \hspace{-0.35cm}$\begin{array}{c} (-\omega^2,1,0)^\mathrm{T} \\[-2pt] +\alpha(-\omega^2,0,1)^\mathrm{T} \end{array}$\hspace{-0.35cm} & \hspace{-0.35cm}$\begin{array}{c} (-\omega^2,1,0)^\mathrm{T} \\[-2pt] +\alpha(-\omega^2,0,1)^\mathrm{T} \end{array}$\hspace{-0.35cm} \\
			\midrule
			$\Z{3}^{(4)}$
			& $\rep{1}\oplus\rep{1}_{\omega}\oplus\rep{1}_{\omega^2}$ & $\rep{1}\oplus\rep{1}_{\omega}\oplus\rep{1}_{\omega^2}$ & $\mathrm{BR(ST)^2}$ &  $(b^*,\omega,a)^\mathrm{T}$
			& $(1,a,b)^\mathrm{T}$ \\
			\midrule
			$\Z2$ & $\rep{1}\oplus\rep{1}_{-1}\oplus\rep{1}_{-1}$ & $\rep{1}\oplus\rep{\rep{1}}\oplus\rep{1}_{-1}$ & $\mathrm{C}$ &  $\begin{array}{c} (0,1,-1)^\mathrm{T} \\[-2pt] \text{(preserves $\Z6$)} \end{array}$ & $\begin{array}{c} (1,0,0)^\mathrm{T} \\[-2pt] +\alpha(0,1,1)^\mathrm{T} \end{array}$\\
			\bottomrule
		\end{tabular}
	}
	\caption{\label{tab:H321}
		Maximal subgroups (up to conjugation) of $H(3,2,1)$ that can be achieved from its breakdown by vevs of flavon 
		fields at $\vev T=\omega$ (similar results hold for $\vev T=1, \I\infty$). We provide the branchings of the flavon 
		representations under the resulting subgroups, 	followed by samples of the generators of such subgroups and the 
		flavon vevs that yield the corresponding breakdown. In the first row of the top block, the subindices in the 
		branching representations correspond to the charges with respect to the \Z3 generated by $(\mathrm{B^2A^2})^2\mathrm{R(ST)^4}$.
		We use the definitions $a:=-1+\eta-\eta^4+\eta^5-\eta^8$ and $b:=-\eta+\eta^4-\eta^8$ with $\eta:=\e^{\nicefrac{2\pi\I}{18}}$.
		Whenever the vev preserves not only the subgroup but a larger symmetry, the resulting symmetry is explicitly given in parentheses.
		In all other aspects, we follow the same notation as in table~\ref{tab:D54breaking}.
	}
\end{table}

In order to clarify whether or not the vevs listed in tables~\ref{tab:Xi22} and~\ref{tab:H321} 
have a \CP-like stabilizer, we take the representative $U$ of the respective \CP-like transformation, 
stated in eqs.~\eqref{eq:CPtrafoT1} and~\eqref{eq:CPtrafoTomega}, and before eq.~\eqref{eq:groupTi}. 
Furthermore, we allow to amend this generator by any element $\rho(\mathrm{g})$ of the unified 
flavor symmetries at the respective moduli location $\vev T$. Then we check whether the flavon vev can solve the equation
\begin{equation}\label{eq:CPlikeStab}
 \rho(\mathrm{g})\,U\,\langle\bar{\Phi}\rangle~\stackrel{!}{=}~\langle\Phi\rangle\;. 
\end{equation}
We find that all of the flavon vevs listed in tables~\ref{tab:Xi22} and~\ref{tab:H321} exhibit $\CP$-like 
stabilizers, with one exception. The sole exception is the vev $(1,1,1)^\mathrm{T}$ for $H(3,2,1)$ 
(as always irrespective of whether $\vev T$ is $\omega$, $1$ or $\I\infty$). 
This vev does not allow for a solution of eq.~\eqref{eq:CPlikeStab} \textit{unless} a rephasing by a 
global phase of $\I$ to $(\I,\I,\I)^\mathrm{T}$ is admitted. Note that also the omission 
of the global prefactors for all other vevs is, of course, an explicit choice of a global phase,
in the sense that eq.~\eqref{eq:CPlikeStab} would be spoiled also for these vevs upon a ``wrong'' 
choice of global phase. Altogether, we see that all of the vevs have \CP-like stabilizers 
(modulo the global choice of phase). In other words, none of the discussed breaking patterns leads itself to \CP violation 
in a model independent way. Nonetheless, \CP will certainly be broken 
(just as the other residual symmetries) once the flavon and/or modulus vevs 
are deflected away from their symmetry enhanced points, or, alternatively, if multiple vevs with 
incommensurable stabilizers are switched on at the same time.
\enlargethispage{\baselineskip}

\section{Conclusions and Outlook}
\label{sec:conclusions}

We have analyzed in detail the breakdown of the eclectic flavor
group as it appears in the top-down approach based on string
theory. In a realistic setup, the eclectic flavor symmetry 
has to be broken. This breaking is induced by two mechanisms: 
First, the vev of the modulus breaks the finite modular symmetry, 
at least partially. Second, the traditional flavor symmetry is 
universal in moduli space and, hence, unbroken by the modulus vev. 
It can be enhanced by elements of the finite modular symmetry at 
specific, symmetry enhanced points in moduli space. 
Thus, in the top-down approach,
one cannot just consider the finite modular flavor
symmetry and ignore the traditional flavor group. The latter
enhances the predictive power of the scheme as it gives severe
restrictions on the K\"ahler potential and superpotential of the theory
(see~\cite[section 3]{Nilles:2020kgo} for a detailed discussion). Of 
course, the (possibly enhanced) traditional
flavor symmetry has to be broken as well, and this requires the
introduction of flavon fields. This increases the number of
parameters, but there is no alternative. The flavon fields that
break the traditional flavor symmetry might break the modular
symmetries as well. This leads to an attractive flavor
structure due to the subtle interplay of the symmetry breakdown
via flavons and moduli as it allows the incorporation of various
possibilities for ``flavor hierarchies'' through the alignment
of vevs.

We illustrate the scheme in detail for an example based on the
$\mathbbm T^2/\Z3$ orbifold sector with traditional flavor group $\Delta(54)$,
modular flavor group $T'$, and eclectic flavor group
$\Omega(2)$ as displayed in table~\ref{tab:Z3FlavorGroups}. For the top-down approach,
we consider a representation content as it appears in orbifold
compactifications of the heterotic string (here, the $\mathbbm T^6/(\Z3\x\Z3)$
orbifold, as discussed in~\cite{Carballo-Perez:2016ooy}). All possible
massless representations are given in table~\ref{tab:Representations}. As usual in the 
top-down approach, the spectrum is very selective, and only a few representations of
$\Delta(54)$ and $T'$ appear as massless modes. In the
present example, we also observe that the automorphy factors
of the fields are strictly correlated with the corresponding
representations of the discrete modular group $T'$. Thus,
there is here no freedom to choose modular weights by hand, they
are fixed by the underlying string theory construction.

As a warm-up example, we discuss the breakdown of $\Delta(54)$
via flavon fields in section~\ref{sec:Delta54Breaking}. The breakdown pattern is shown
in figure~\ref{fig:D54breaking} and table~\ref{tab:D54breaking}. The main result of the paper concerns
the breakdown patterns of the eclectic flavor group $\Omega(2)$, which 
is derived in section~\ref{sec:Omega2Breaking}. We specifically consider the flavor groups
$\Xi(2,2)=[324,111]$ and $H(3,2,1)=[486,125]$, which appear as
unbroken subgroups of $\Omega(2)$ at the fixed points $T=\I$ and
$T=1, \omega$ (as well as at $T=\I\infty$ which is dual to $T=1$), respectively. The qualitative breakdown patterns
via flavon fields are summarized in figures~\ref{fig:Xi22breaking} and~\ref{fig:H321breaking}. 
The specific form of the corresponding flavon vevs is given in
tables~\ref{tab:Xi22} and~\ref{tab:H321}. This shows that even a simple system like
the $\mathbbm T^2/\Z3$ orbifold sector exhibits a rich web of breakdown patterns
via flavon and modulus vevs that might be suitable to be applied
to discuss the flavor structure of quarks and leptons in the
standard model of particle physics. In a companion paper~\cite{Baur:2021pr}, 
we shall show that a successful fit of the masses and
mixing angles of quarks and leptons can be achieved.

\subsection*{Acknowledgments}
P.V. and A.B. were supported by the Deutsche Forschungsgemeinschaft (SFB1258).
H.P.N. thanks the ASC-LMU and in particular Dieter L\"ust for hospitality and support.

\providecommand{\bysame}{\leavevmode\hbox to3em{\hrulefill}\thinspace}

\end{document}